\documentclass[final,3p,times]{elsarticle}

\usepackage{multirow,setspace,times,amssymb,amsmath,graphicx,color,rotating,subfigure,url}
\usepackage{lineno}
\usepackage{natbib}

\usepackage{epsf}
\usepackage{rotating}

\usepackage{graphicx,epsfig}
\usepackage{dcolumn}
\usepackage{bm}

\journal{Commun. Nonlinear Sci. Numer. Simulat.}

\begin{document}

\title{Information spreading on dynamic social networks}

\author[inst1]{Chuang Liu}
\ead{liuchuang@hznu.edu.cn}
\author[inst1]{Zi-Ke Zhang}
\ead{zhangzike@gmail.com}

\address[inst1]{Institute of Information Economy, Hangzhou Normal University, Hangzhou 310036, P.R. China}

\begin{abstract}
Nowadays, information spreading on social networks has triggered an
explosive attention in various disciplines. Most of previous
works in this area mainly focus on discussing the effects of
spreading probability or immunization strategy on static networks.
However, in real systems, the peer-to-peer network structure changes
constantly according to frequently social activities of users. In
order to capture this dynamical property and study its impact on
information spreading, in this paper, a link rewiring strategy based
on the Fermi function is introduced. In the present model, the
informed individuals tend to break old links and reconnect to their second-order friends
with more uninformed neighbors. Simulation results on the
susceptible-infected-recovered (\textit{SIR}) model with fixed
recovery time $T=1$ indicate that the information would spread more faster
and broader with the proposed rewiring strategy. Extensive analyses of the
information cascade size distribution show that the spreading process of the initial
steps plays a very important role, that is to say, the information
will spread out if it is still survival at the beginning time. The
proposed model may shed some light on the in-depth understanding of
information spreading on dynamical social networks.
\end{abstract}

\hspace{4pc}{\bf Keywords}: information spreading, social networks,
dynamical rewiring

\maketitle

\section{\label{S1:Intro}Introduction}

The epidemic spreading based on complex networks, where nodes
represent individuals or organizations and links denote their
interactions, has attracted an increasing attention in recent years
 \cite{Castellano-Fortunato-Loreto-2009-RMP,Lloyd-May-2001-Science,Newman-2002-PRE}.
Epidemic spreading is a dynamic process in which an item is
transmitted from an infected individual to a susceptible individual
through the link between them. Therefore, the network structure is a
particularly important factor for the efficiency of epidemic
spreading. Recently, many pioneering works about
susceptible-infected-susceptible ($SIS$) and
susceptible-infected-recovered ($SIR$) models indicate that a highly
heterogeneous structure will lead to the absence of any outbreak
threshold \cite{Pastor-Satorras-Vespignani-2001-PRL} while the
epidemic spreading on small-world network exhibits critical behavior
 \cite{Zanette-2001-PRE}. The voluntary vaccination strategy under
game theory framework shows that the epidemic spreading on
scale-free networks can be favorably and easily
controlled \cite{Zhang-Zhang-Zhou-Small-Wang-2010-NJP, wang2012global}. However, all
those interesting results are obtained based on the research of the
static network, where interactions are always fixed. By contrast, in real
online systems, people communicate with various individuals and might make
new friends everyday. That is to say, the social communication
network, also referred to as the peer-to-peer network,  would change
its topology dynamically. Consequently, it would be very suitable to study
such dynamic networks with the \textit{rewiring
strategy} \cite{Holme-Newman-2006-PRE}, where the network structure
changes by breaking old links and forming new ones.

In the past few years, many researches
have focused on the epidemic spreading problem in such dynamically
contacting networks based on the link rewiring
strategy \cite{Kossinets-Watts-2006-Science,Gross-Dlima-Blasius-2006-PRL,Schwartz-Shaw-2010-Physics,Shaw-Schwartz-2008-PRE,Wieland-Aquino-Nunes-2012-EPL,Yoo-Lee-Kahng-2011-PA,Volz-Meyers-2007-PRSB}.
The most important and widely used one is the \textit{adaptive
model} \cite{Gross-Dlima-Blasius-2006-PRL,Schwartz-Shaw-2010-Physics},
in which the susceptible individuals try to avoid contacting
the infected
individuals \cite{Gross-Dlima-Blasius-2006-PRL,Shaw-Schwartz-2008-PRE,Wieland-Aquino-Nunes-2012-EPL}.
Simulation results of $SIS$
 \cite{Gross-Dlima-Blasius-2006-PRL,Wieland-Aquino-Nunes-2012-EPL}
and susceptible-infected-recovery-susceptible ($SIRS$)
models \cite{Shaw-Schwartz-2008-PRE} on adaptive networks show that
the epidemic outbreak threshold would be larger than that on static networks.
It indicates that the rewiring process typically tends to suppress
epidemic spreading via isolating infected individuals.
Recently, Yoo \textit{et al} \cite{Yoo-Lee-Kahng-2011-PA} proposed
a \emph{fitness-adaptive rewiring} model where each individual's
degree is preserved in the adaptive model
 \cite{Volz-Meyers-2007-PRSB}. They found that the speed of approaching the epidemic threshold is
delayed and the prevalence is reduced comparing with adaptive
models. Above all, those epidemic spreading researches on dynamic networks
based on the adaptive model indicate that segregating infected
individuals (or susceptible individuals) is an efficient strategy of
reducing the fraction of susceptible-infected interactions, as well as
preventing the outbreak of the whole spreading process.

\begin{figure}[htb]
  \centering
  \includegraphics[width=10cm]{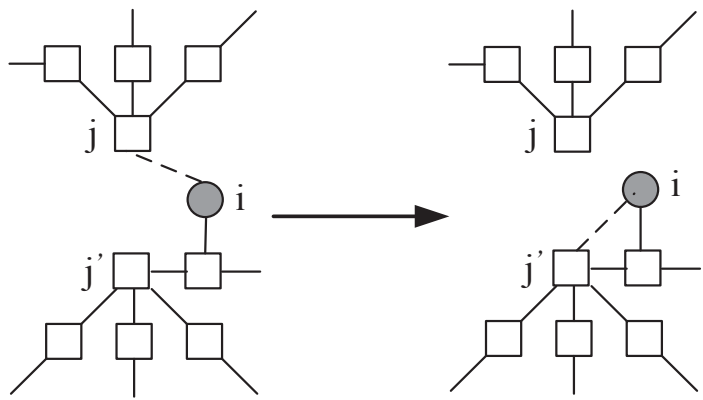}
  \caption{(Color online)\label{Fig:Model} Illustration of the one-step rewiring
  process for a given node $i$. The left panel is the original network, and the
  gray circle $i$ represents the informed individual ($I$-state), while squares represent the
  uninformed individuals ($S$-state). In the original network, individual $j$ is $i$'s neighbor and $j'$ is
  $i$'s second-order neighbor. The payoffs of $j$ and $j'$ are $\pi_j=3$ and $\pi_{j'}=4$ respectively.
  Node $i$ will break the link to $j$ (left panel) and reconnect to $j'$ (right panel) with probability
  $p_f=\frac{1}{1+e^{-\beta(\pi_{j'}-\pi_{j})}}$.}
\end{figure}

However, information spreading is quite different from
disease infections due to its specific features, such as time
decaying effect \cite{Wu-Huberman-2007-PNAS}, tie
strength \cite{Miritello-Moro-Lara-2011-PRE}, information
contents \cite{Crane-Sornette-2008-PNAS}, memory
effects \cite{Dodds-Watts-2004-PRL}, social
reinforcement \cite{Medo-Zhang-Zhou-2009-EPL,Cimini-Medo-Zhou-Wei-Zhang-2011-EPJB},
non-redundancy of contacts \cite{Lu-Chen-Zhou-2011-NJP}, etc. In this
paper, we propose a new rewiring model to study information
spreading on dynamic networks where individuals will select the
neighbors with larger payoff \cite{Graser-Xu-Hui-2009-EPL} following
the Fermi function \cite{Szabo-Toke-1998-PRE,Traulsen-Pacheco-Nowak-2007-JTB,Fu-Rosenbloom-Wang-Nowak-2011-PRSB}.
In conventional statistical physics, Fermi function is used to describe the probability of occupancy for an electron energy state at certain energy level
by an electron. In the present model, we consider such energy level as payoff of rewiring strategy. That is to say, a given node will change its connection by comparing
the alternatives' payoff. There is already a vast class of researches trying to apply Fermi function in modeling social dynamics. Fu \emph{et al.} \cite{Fu-Rosenbloom-Wang-Nowak-2011-PRSB} used the Fermi function to evaluate the expected costs and benefits of vaccination via
exploring the roles of individual imitation behaviour and population structure. Zhang \emph{et al.} \cite{Zhang-Zhang-Weissing-Perc-Xie-Wang-2013-PLoSONE} considered that individual would adopt rewiring or migration reaction to adverse neighborhoods following the Fermi function and the mixture of different reactions led to much more favorable for the evolution of cooperation. Pacheco \emph{et al.} \cite{Pacheco-Traulsen-Nowak-2006-PRL}
adopted the pair-wise comparison strategy for seeking new interactions of rational individuals based on Fermi function. Analogously, Santos \emph{et al.} \cite{Santos-Pacheco-Lenaerts-2006-PCB}
proposed a computational model to allow individuals to be able to self-organize their social ties, based exclusively on their self-interest,
in order to solve the evolutionary cooperative-competitive dilemma. Van Segbroeck \emph{et al.} \cite{VanS-Santos-Now-Lenaerts-2008-BMCEB} alternatively introduced a model
using the payoff-dependent Fermi function. They demonstrated that defectors were more rapid to break adverse links in order to achieve maximum fitness state,
which finally led to a more heterogeneous network structure and improved cooperators' survivability.

Different from adaptive models, in this model, the informed
individual will break the susceptible-infected link if the
susceptible individual's payoff (the number of susceptible neighbors of the
considering individual) is less than one randomly selected among its second-order neighbors, and
rewire the link to the selected susceptible individual (see Fig.~\ref{Fig:Model}). Simulation results on various networks show that
the spreading on dynamic networks is more efficient than that on static
networks. Especially for the scale-free network, the information
spreading prevalence forms two regimes, indicating that the information
diffusion either dies out quickly or spreads into a finite fraction of
the total population.

\section{Model}

\begin{figure}[htb]
  \centering
  \includegraphics[width=7.5cm]{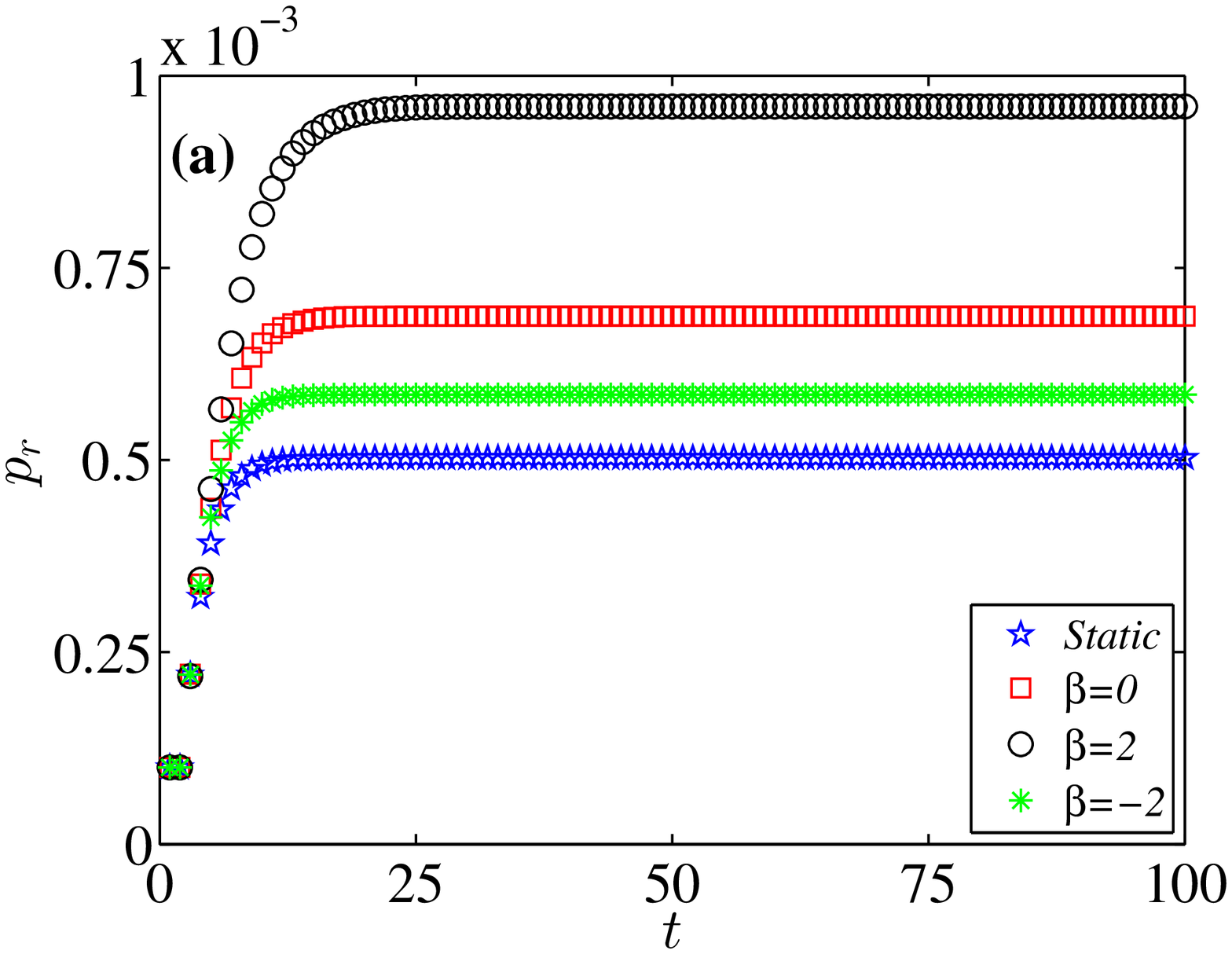}
  \includegraphics[width=7.5cm]{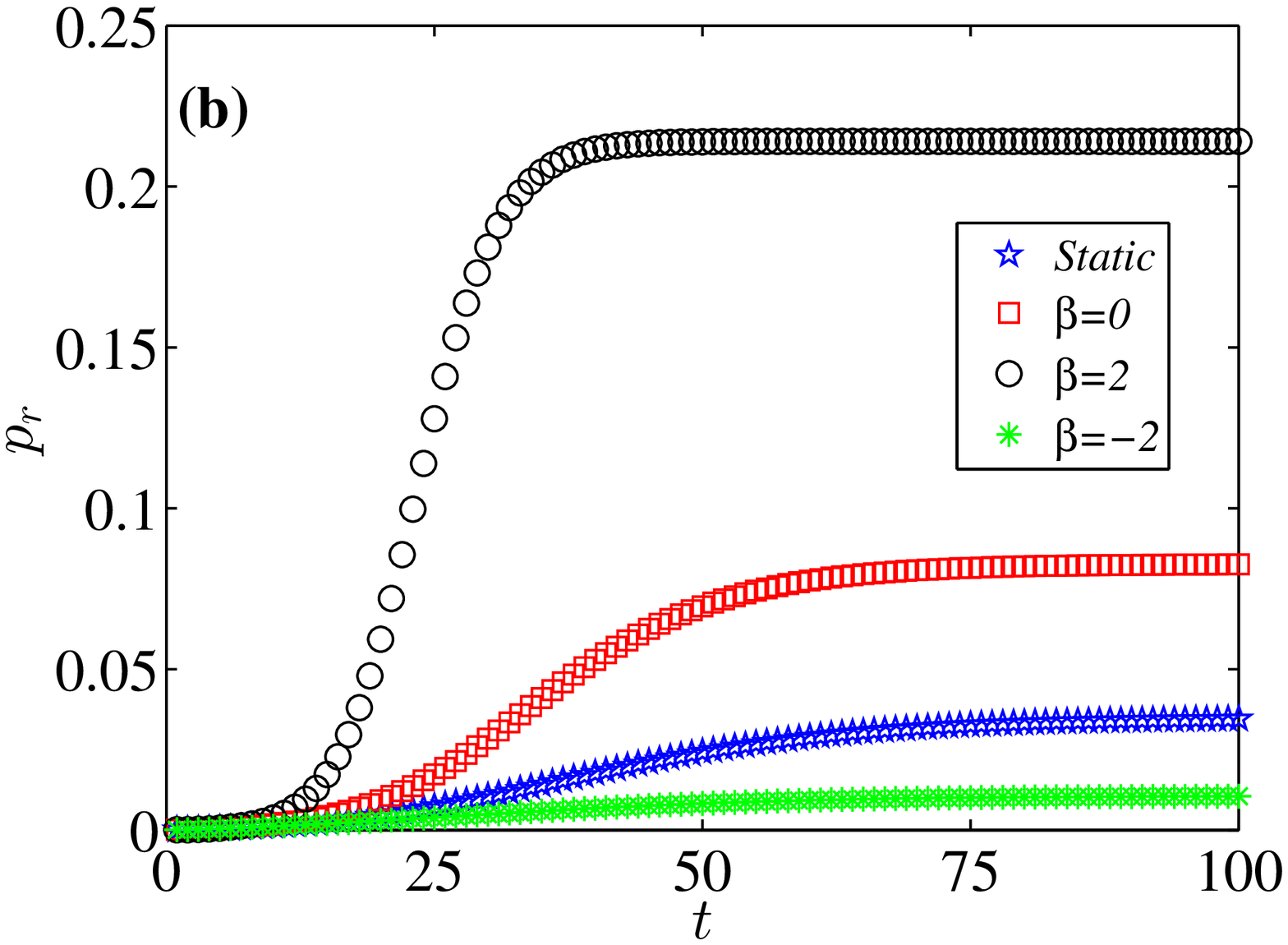}
  \includegraphics[width=7.5cm]{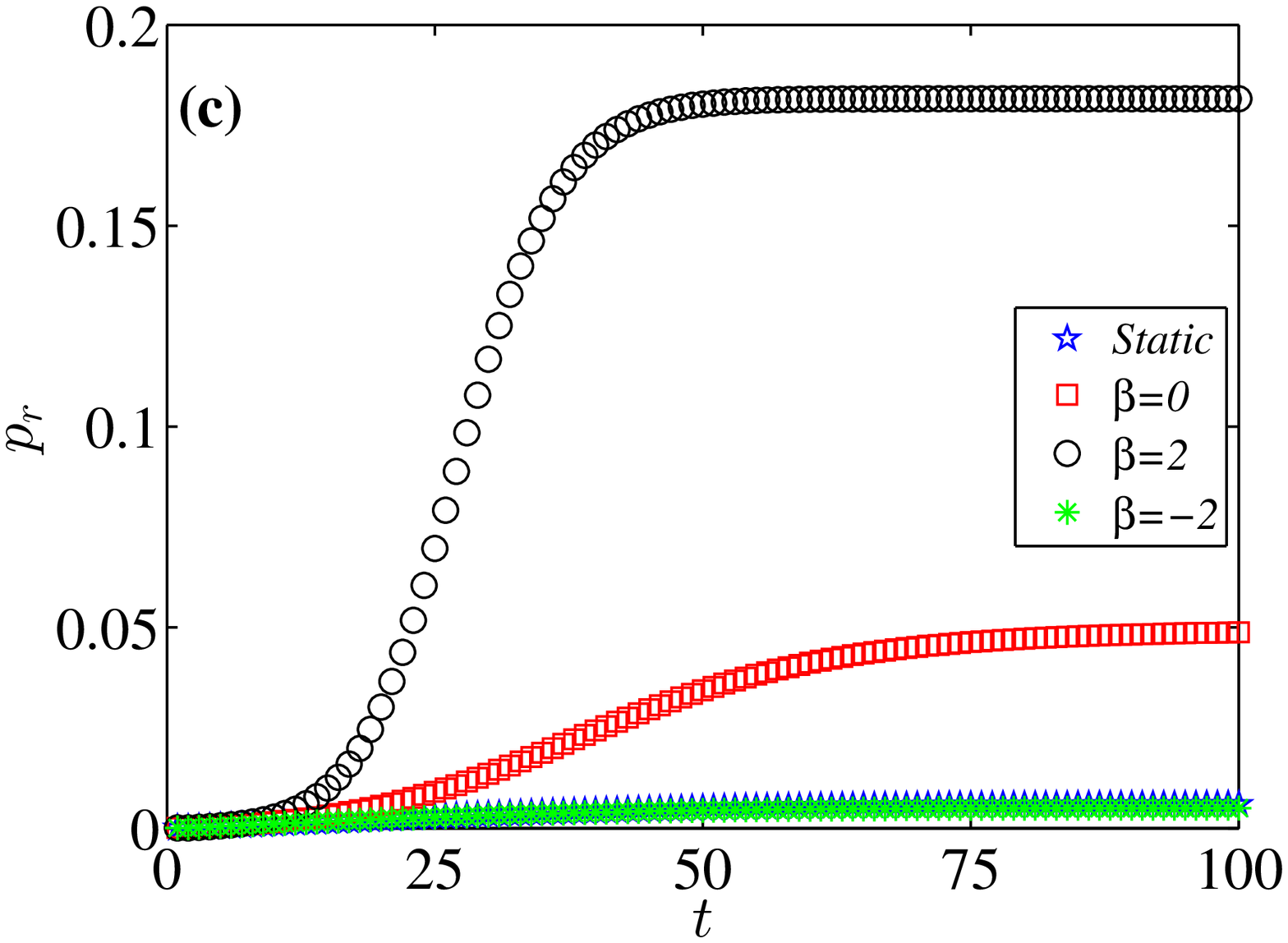}
  \includegraphics[width=7.5cm]{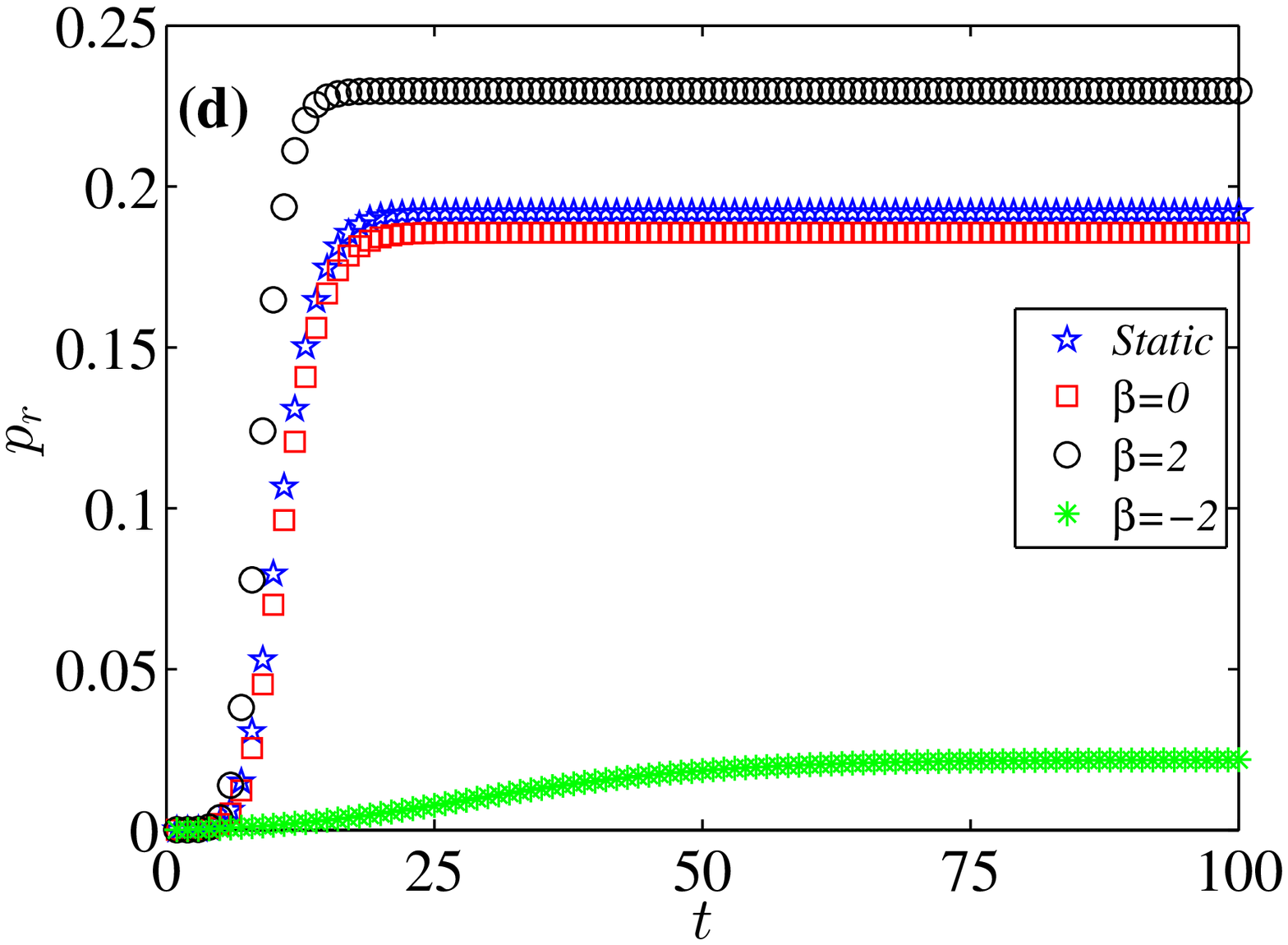}
  \caption{(Color online)\label{Fig:P_time_lambda} Dynamics of $p_r$ with different methods:
   static network (blue pentagram), $\beta=0$ (red square), $\beta=2$
   (black circle) and $\beta=-2$ (green star). All simulations are run on four representative networks: (a) Regular network; (b) Random
  network; (c) Small-world network; (d) Scale-free network. The spreading rate is set to $\lambda=0.2$. }
\end{figure}

\begin{figure*}[htb]
  \centering
  \includegraphics[width=4cm]{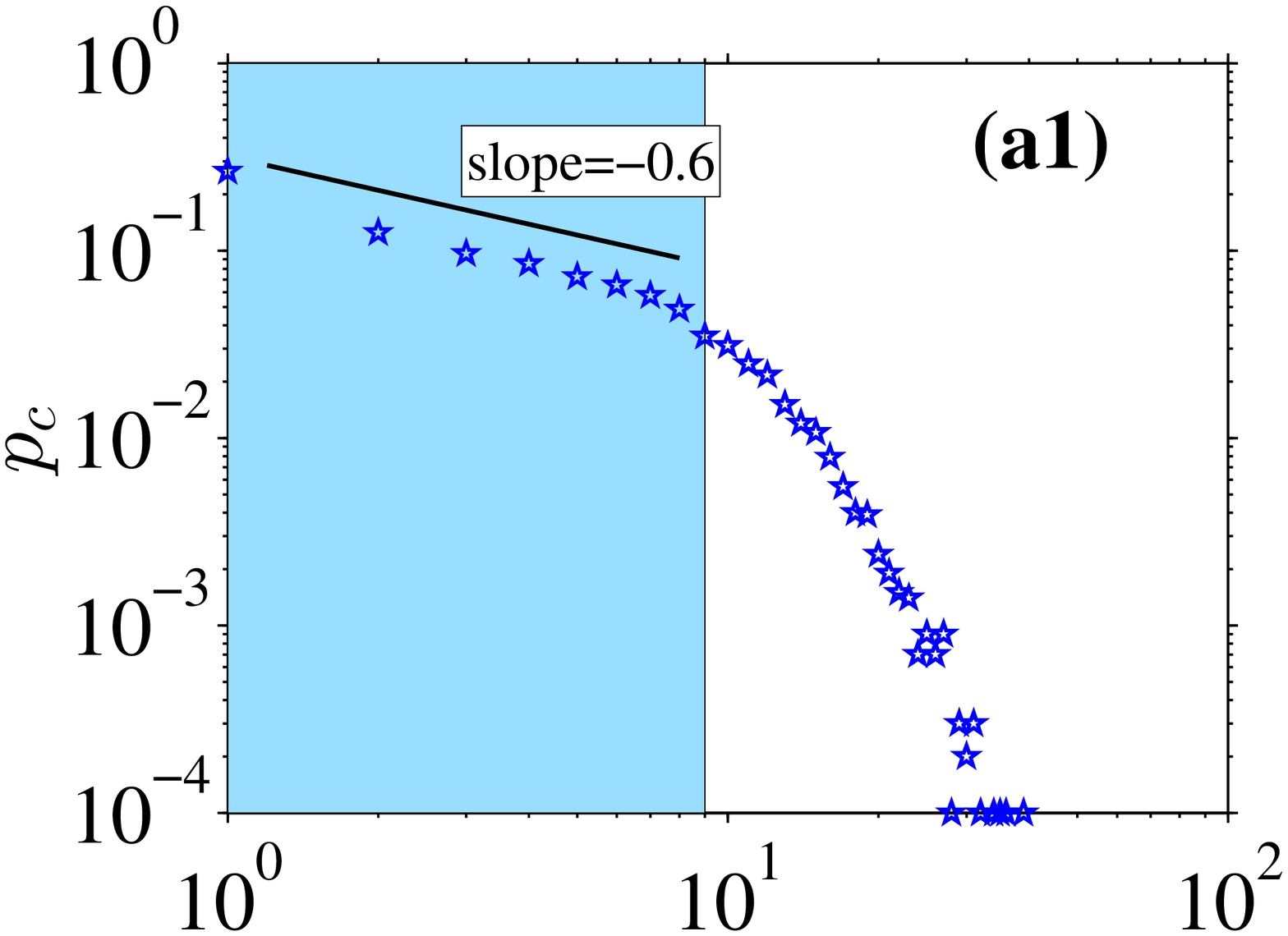}
  \includegraphics[width=3.8cm]{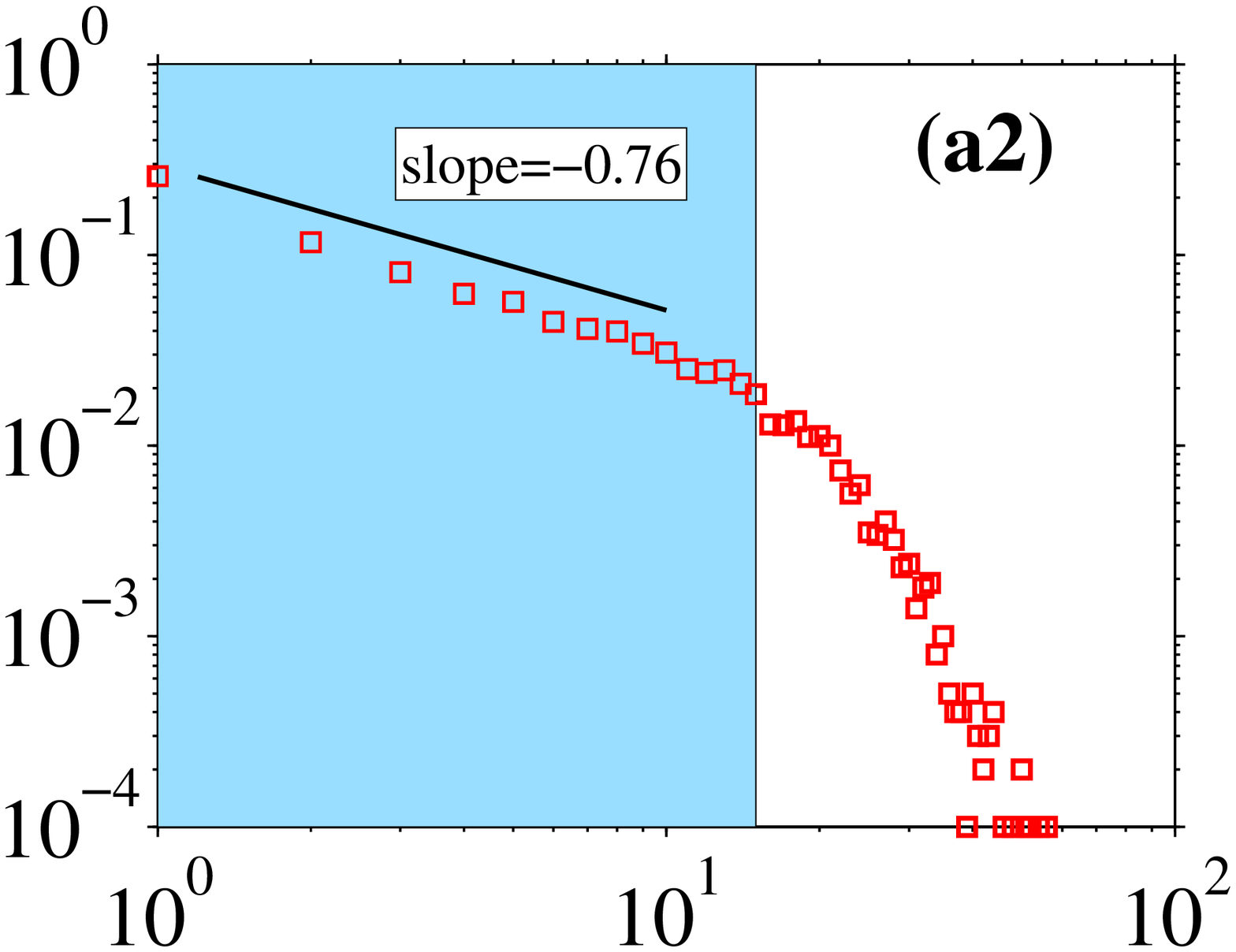}
  \includegraphics[width=3.8cm]{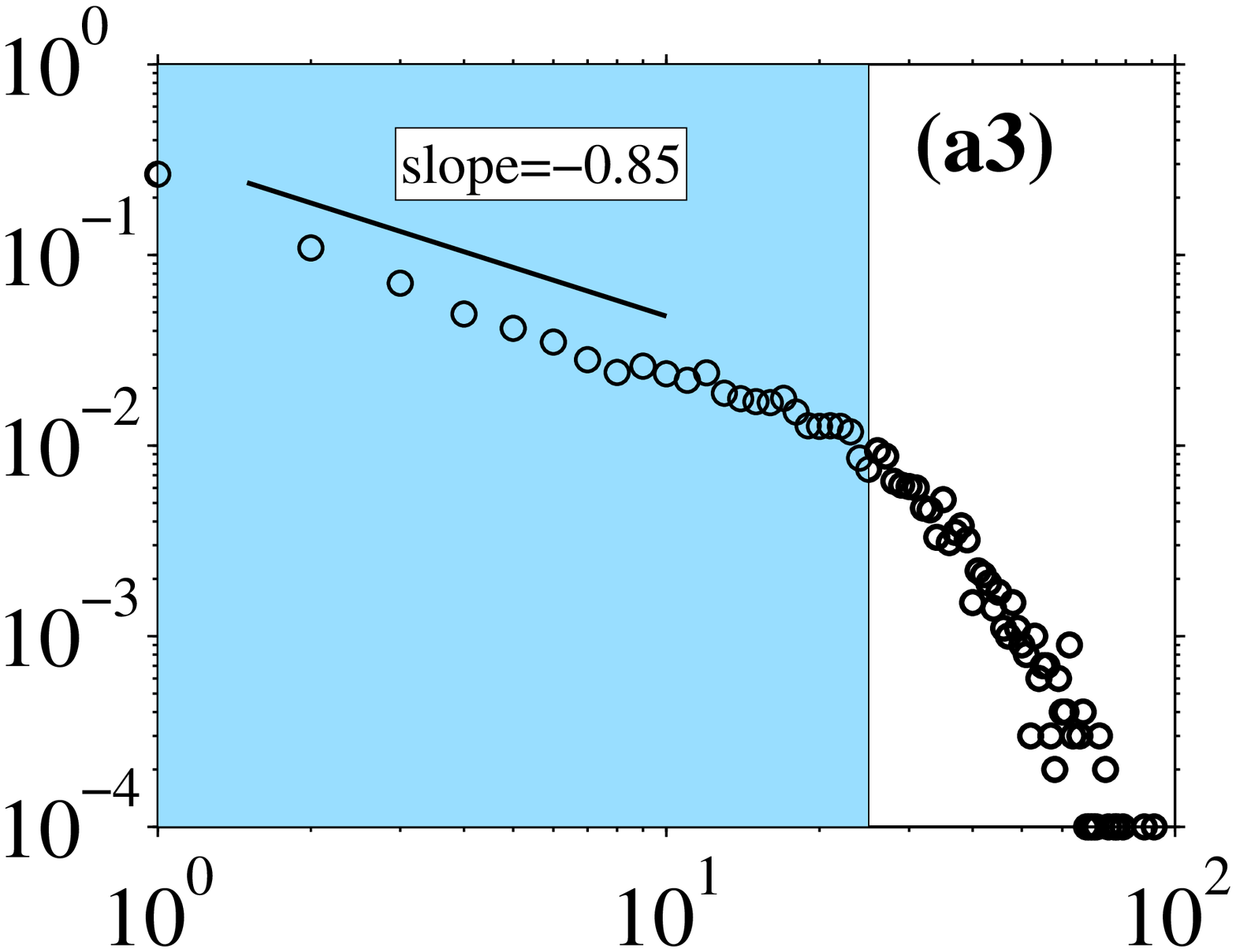}
  \includegraphics[width=3.8cm]{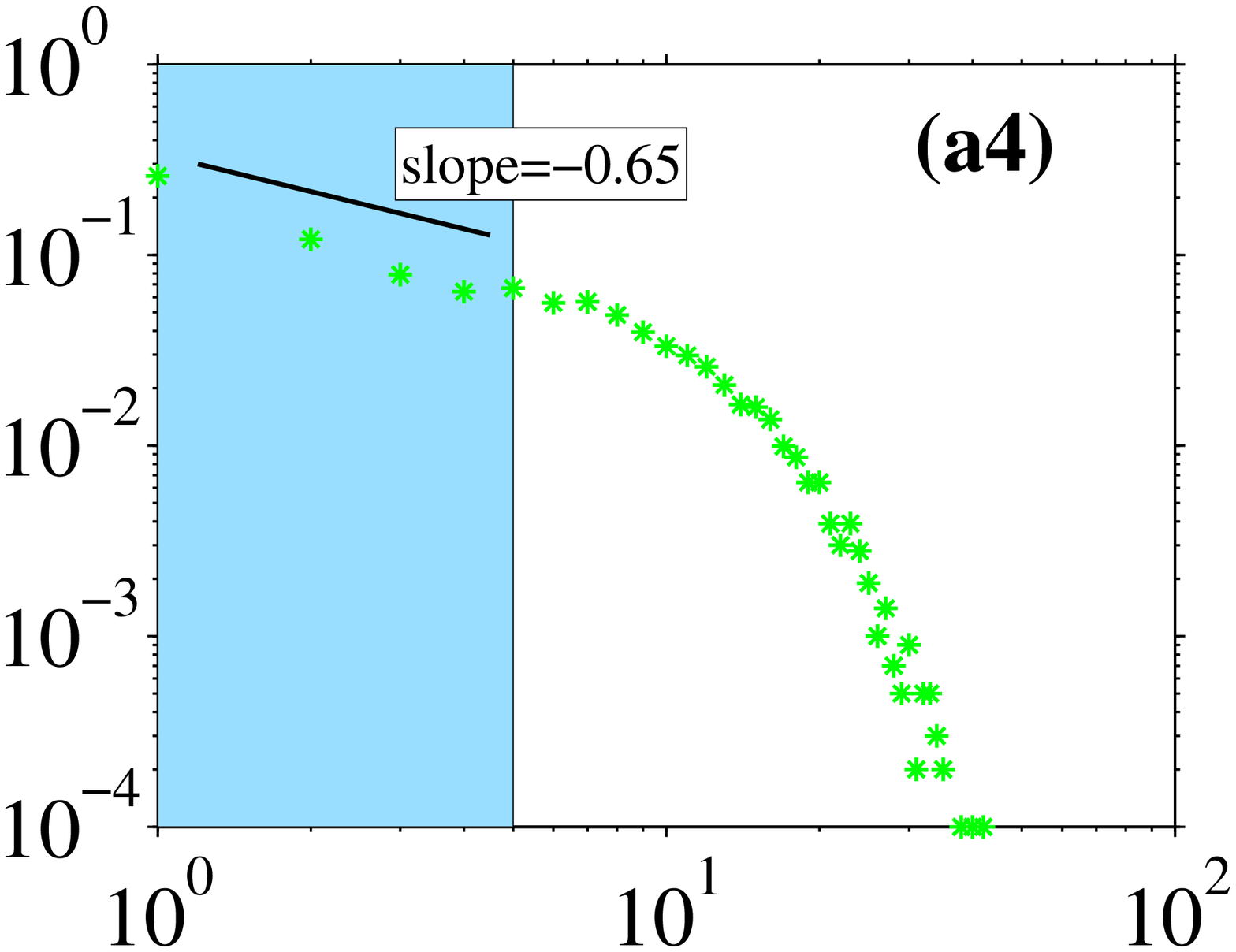}
  \includegraphics[width=4cm]{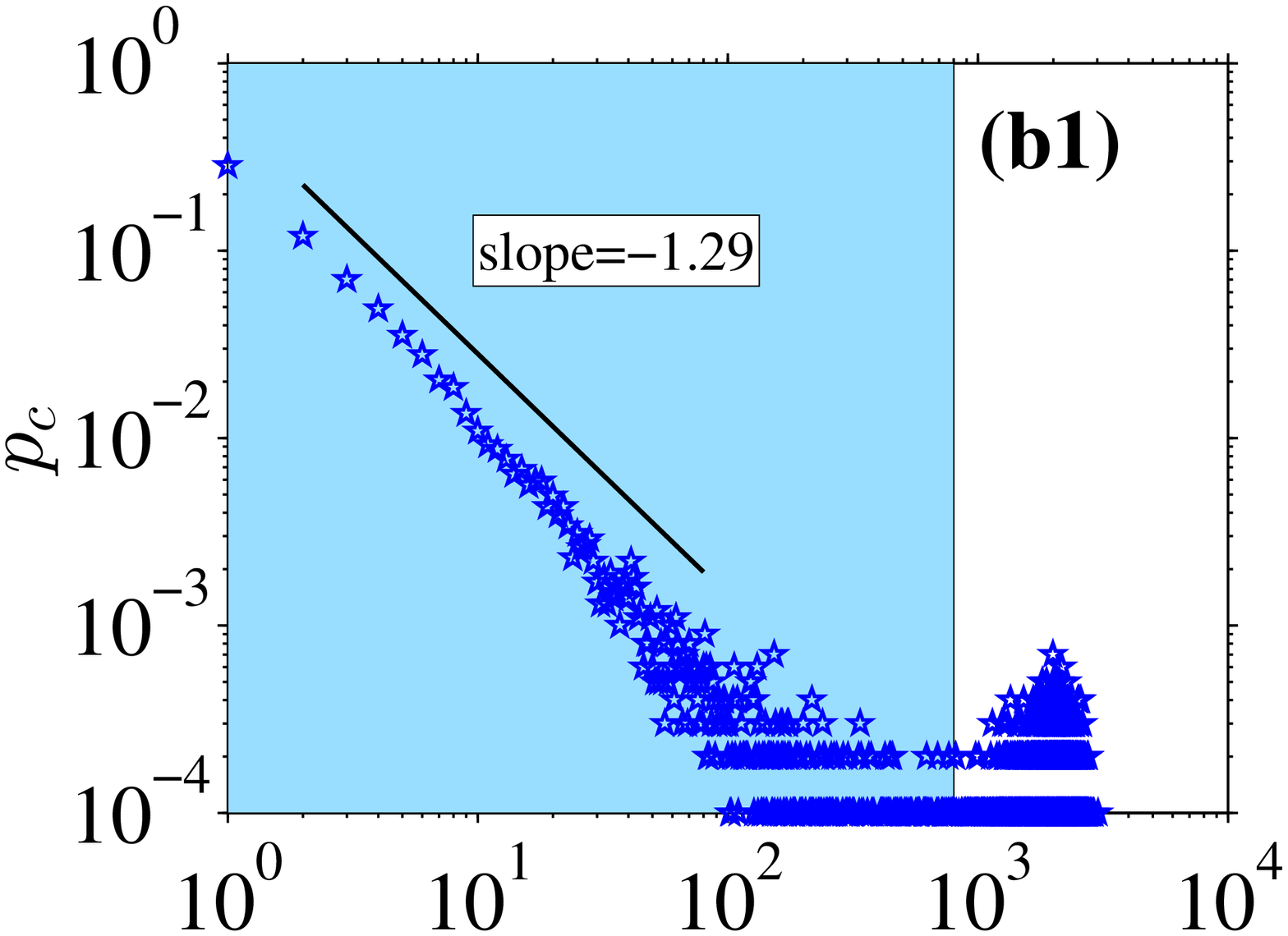}
  \includegraphics[width=3.8cm]{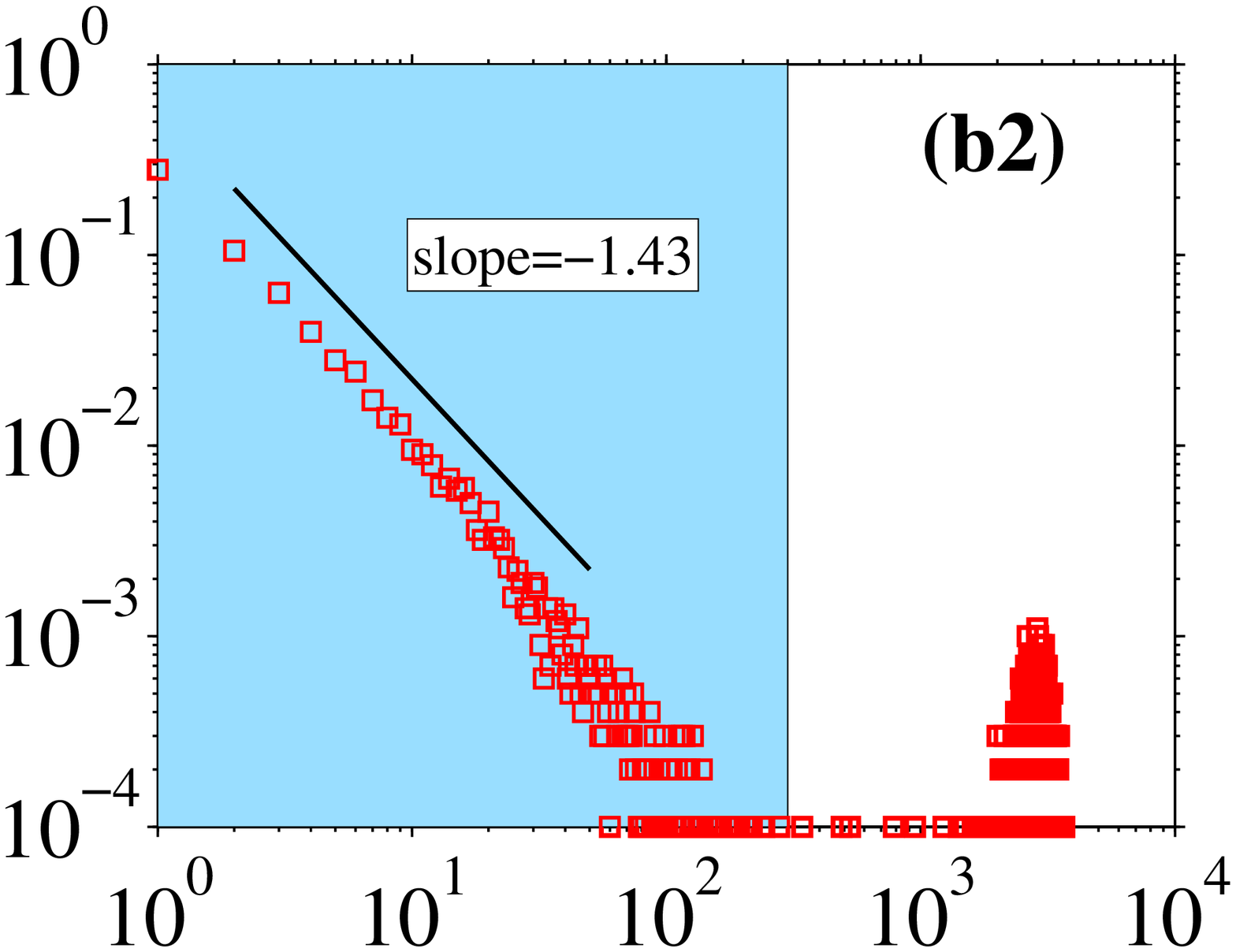}
  \includegraphics[width=3.8cm]{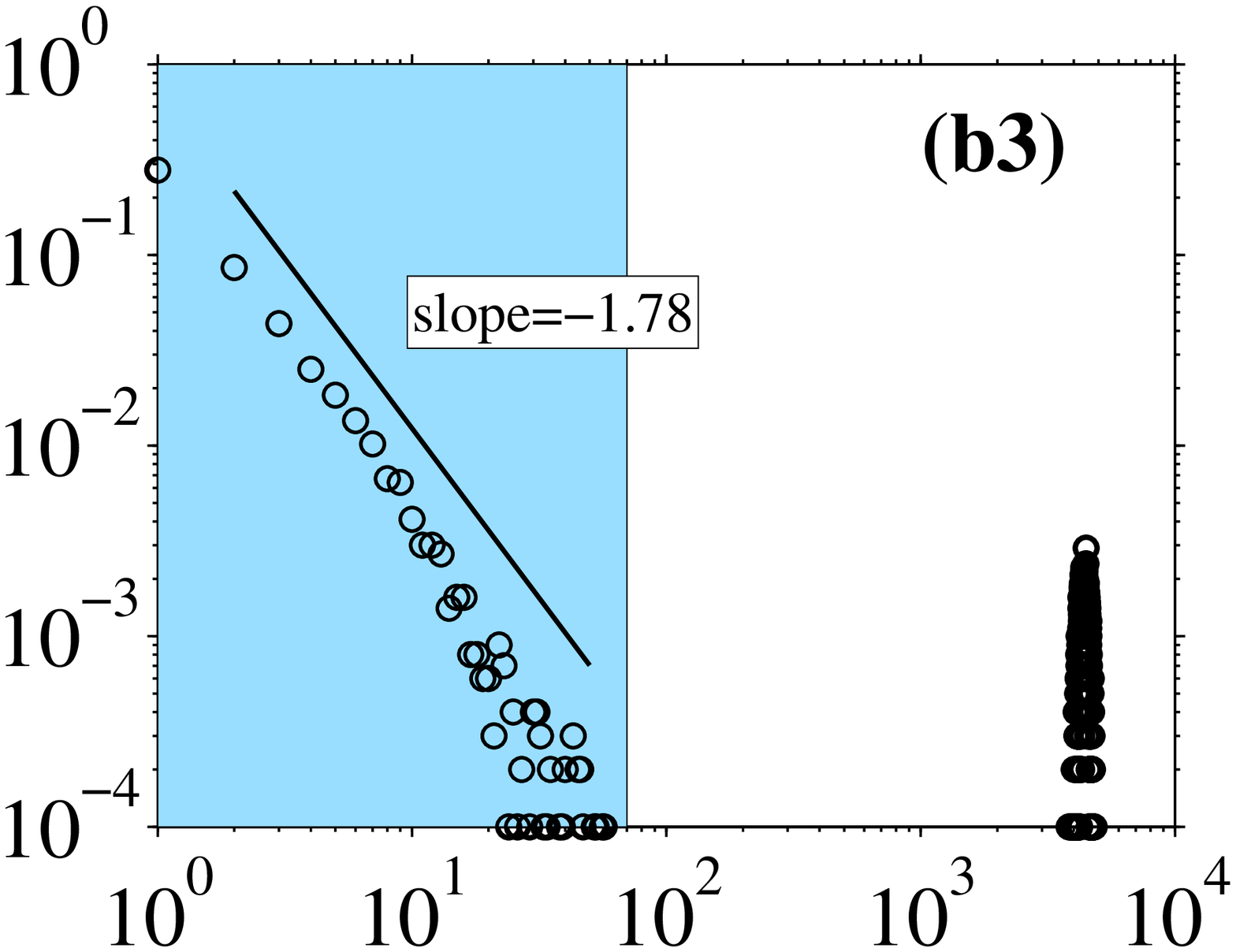}
  \includegraphics[width=3.8cm]{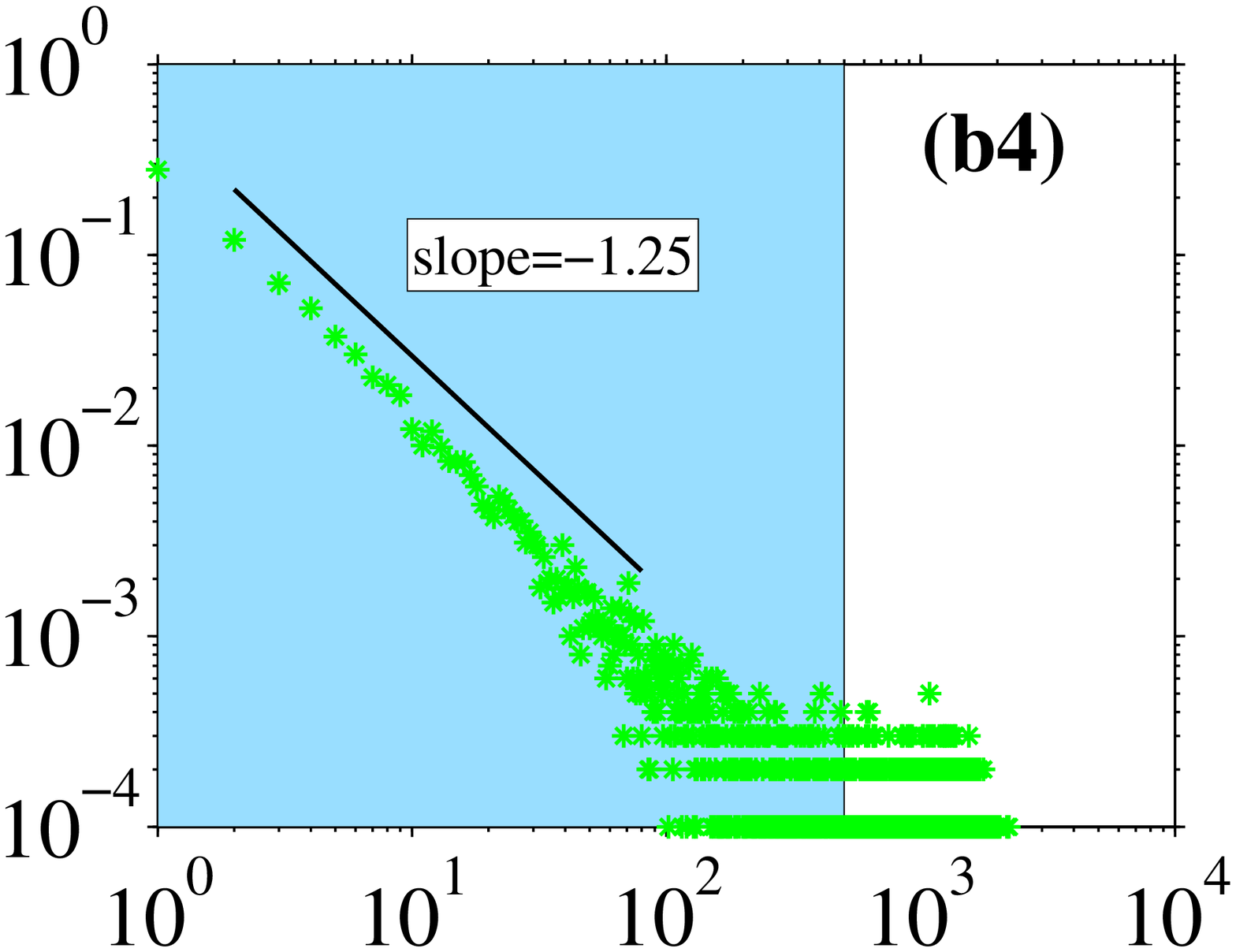}
  \includegraphics[width=4cm]{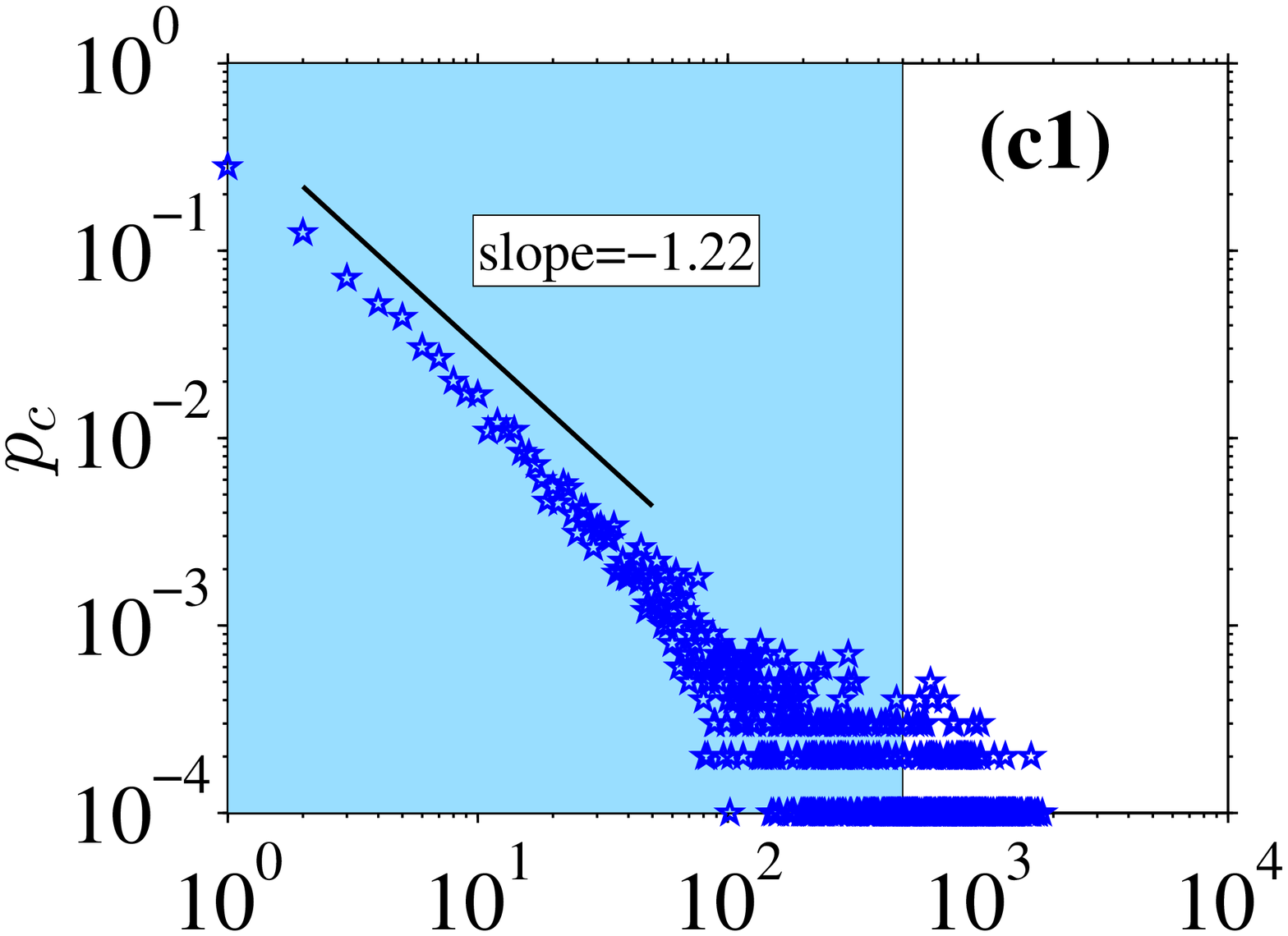}
  \includegraphics[width=3.8cm]{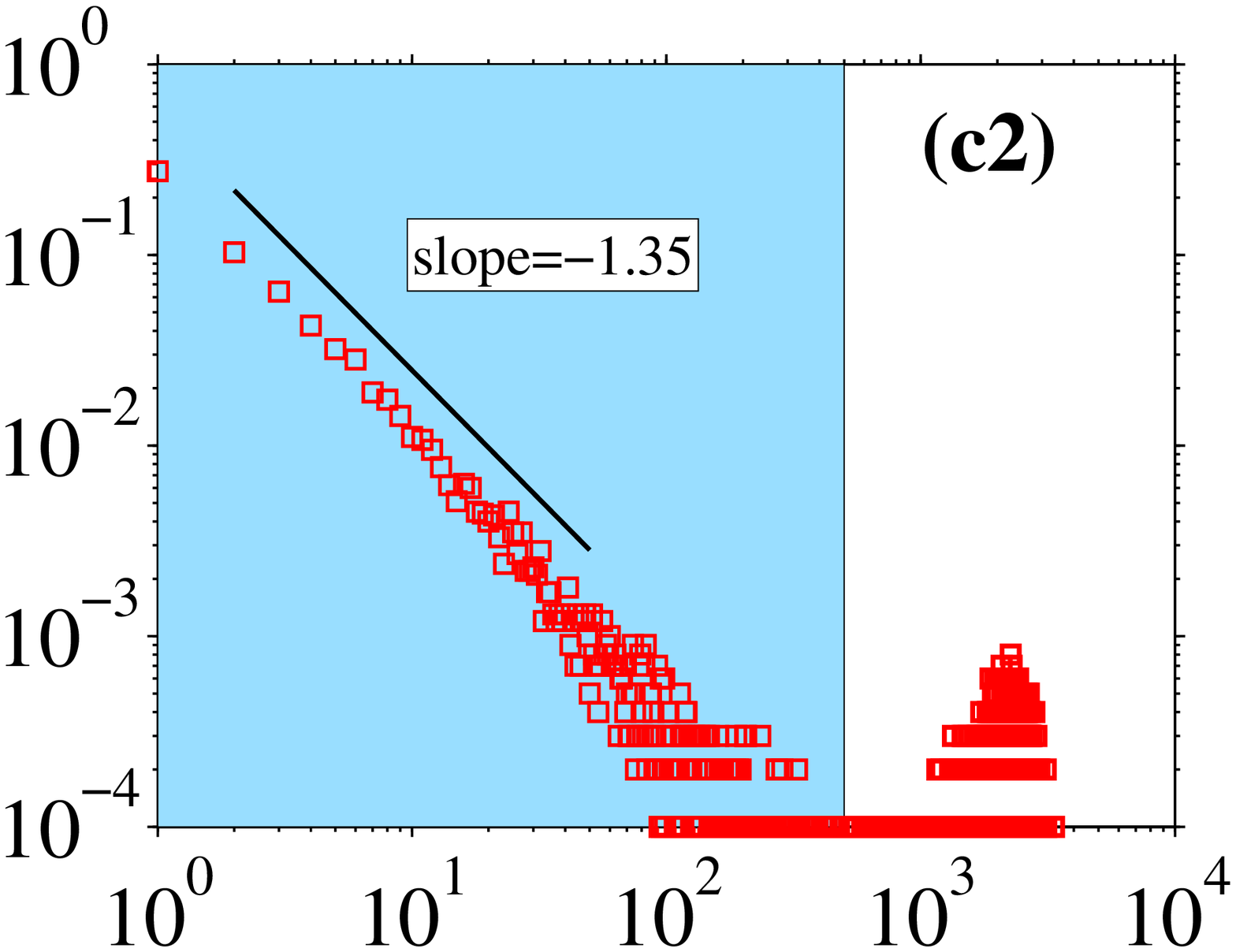}
  \includegraphics[width=3.8cm]{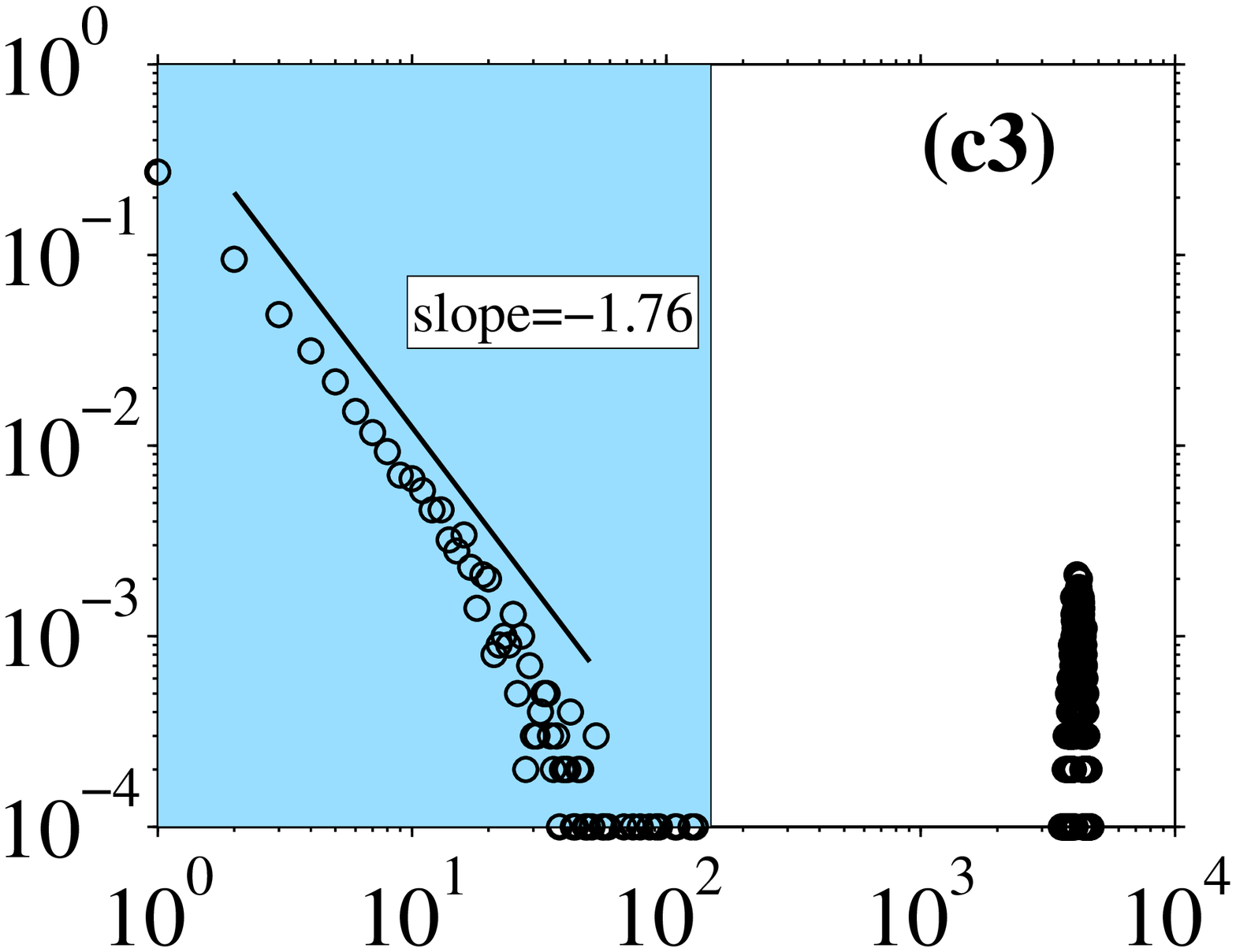}
  \includegraphics[width=3.8cm]{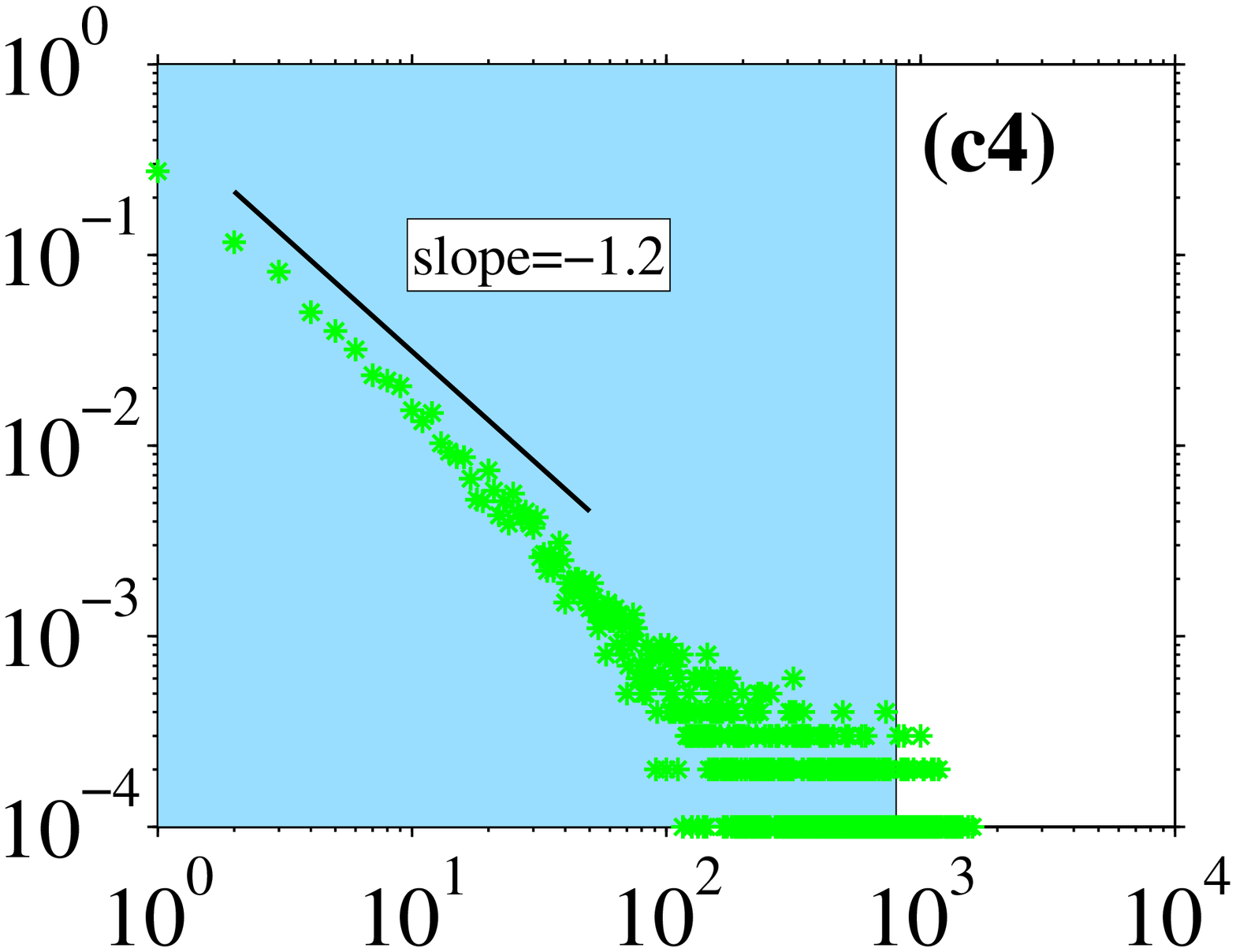}
  \includegraphics[width=4cm]{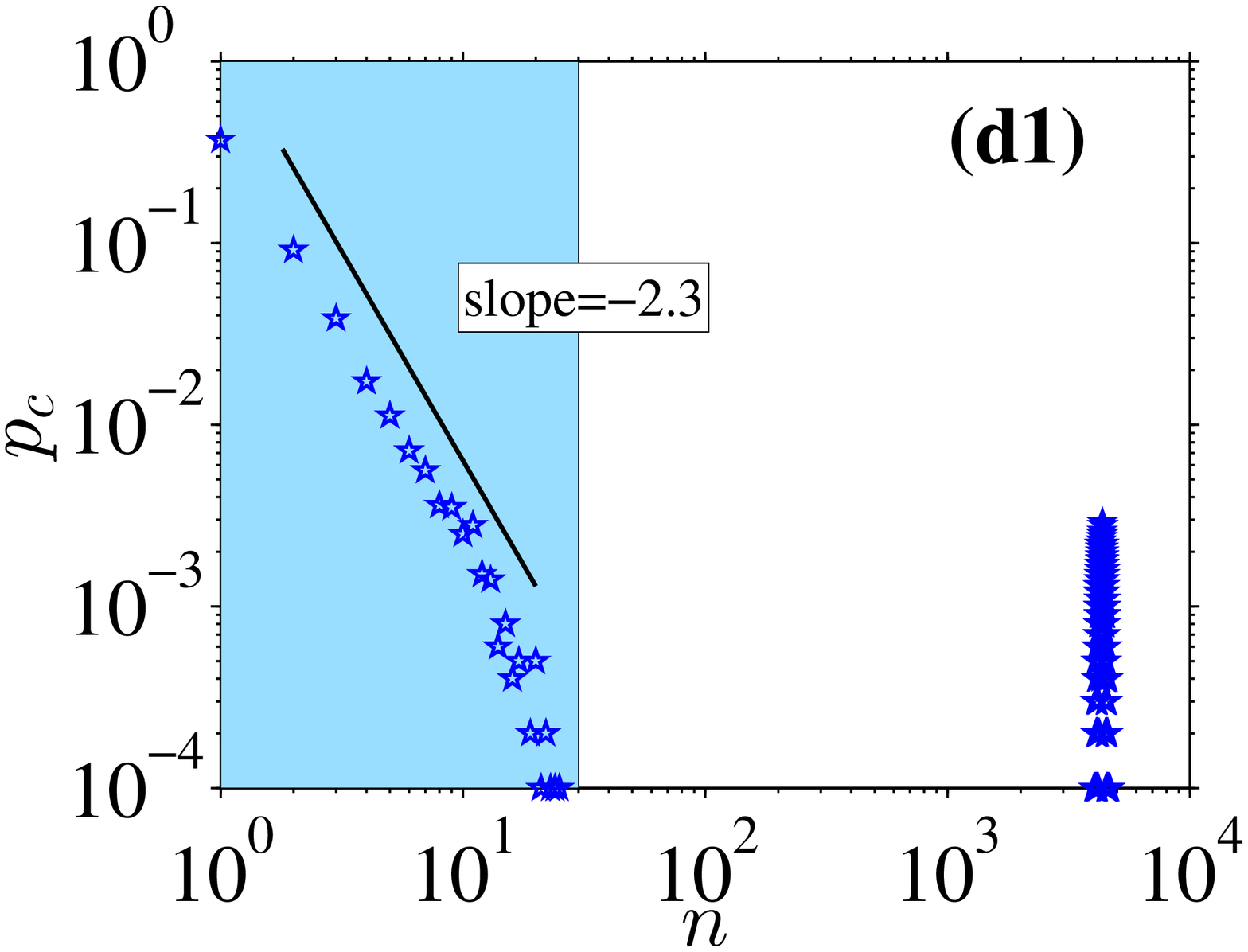}
  \includegraphics[width=3.8cm]{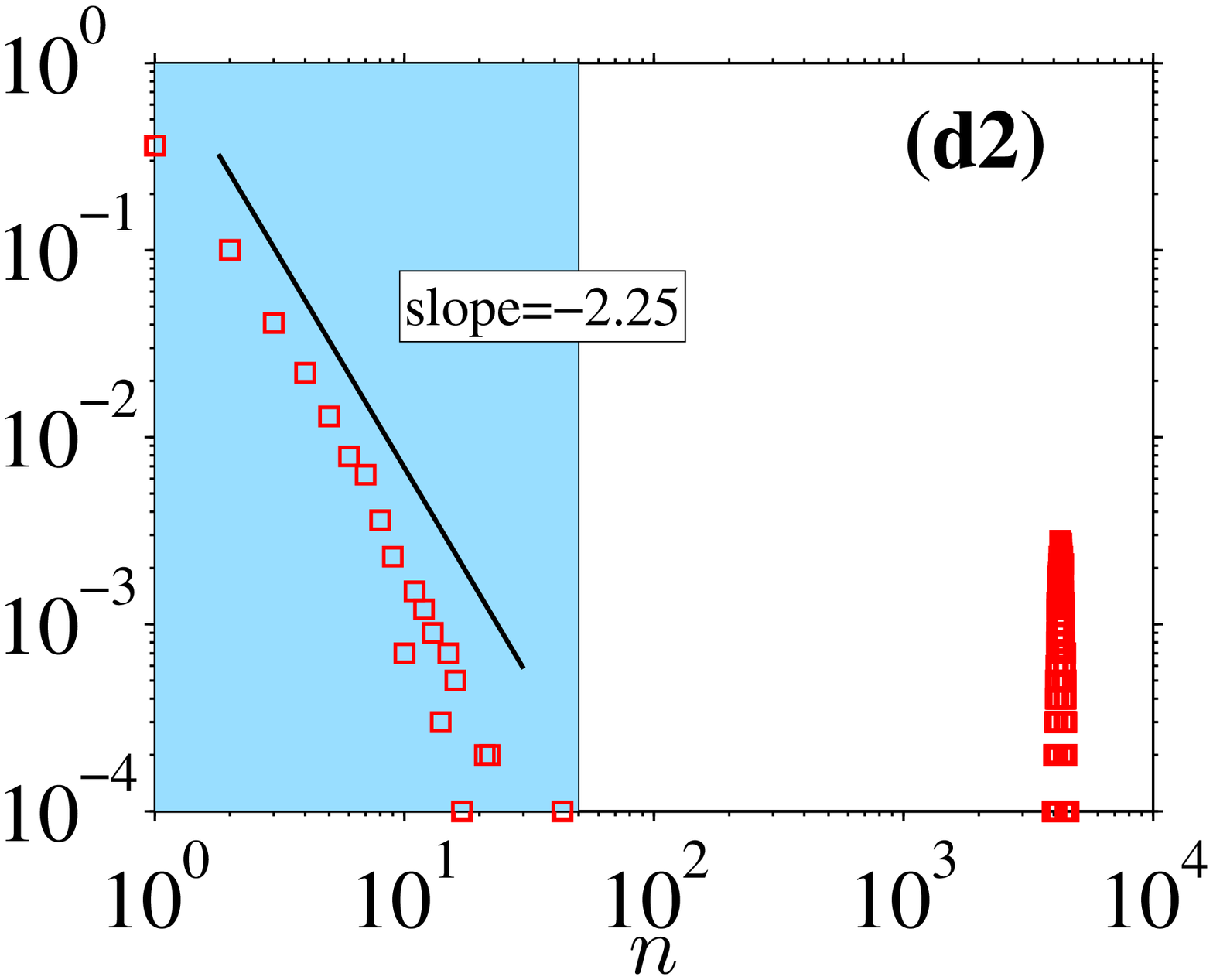}
  \includegraphics[width=3.8cm]{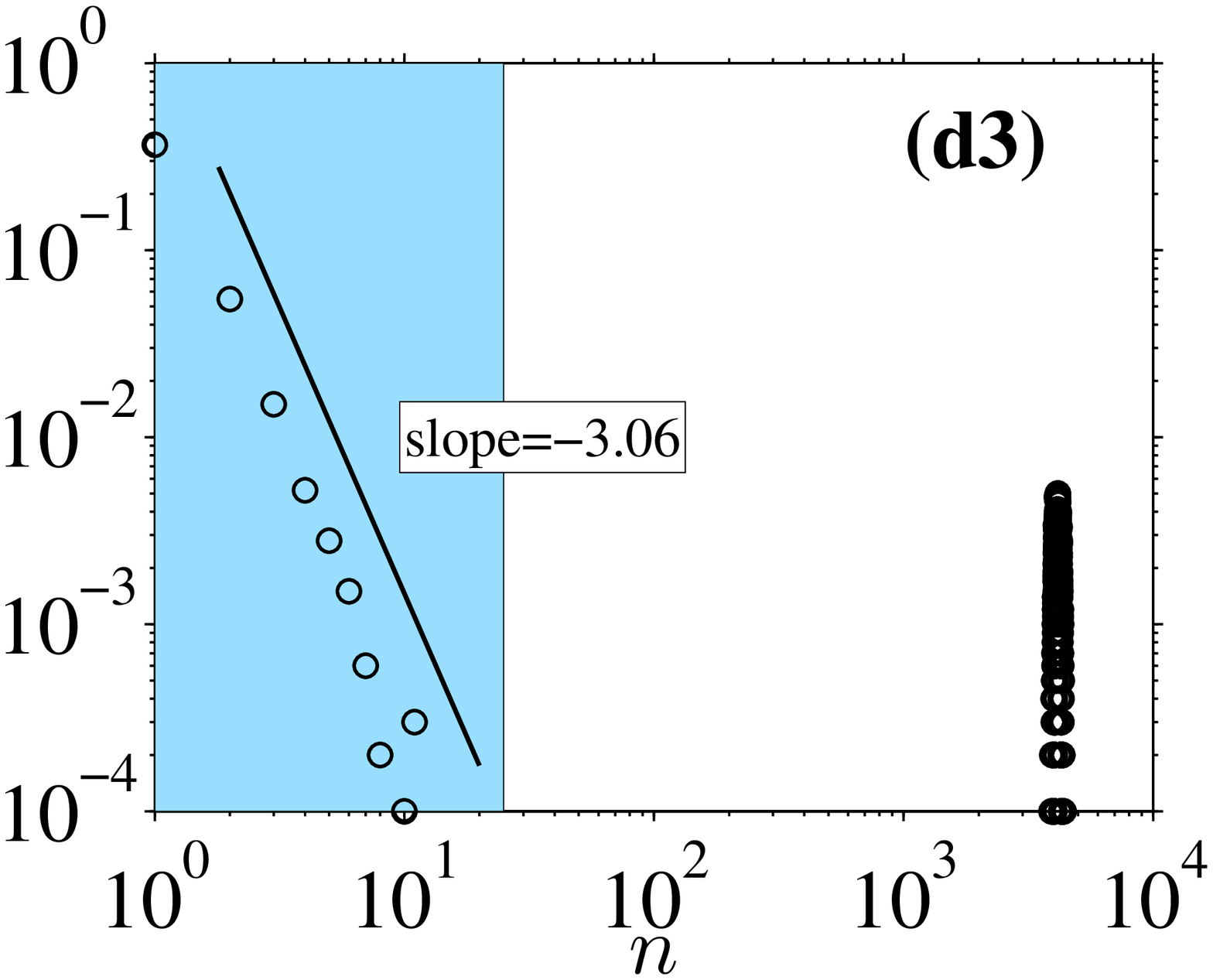}
  \includegraphics[width=3.8cm]{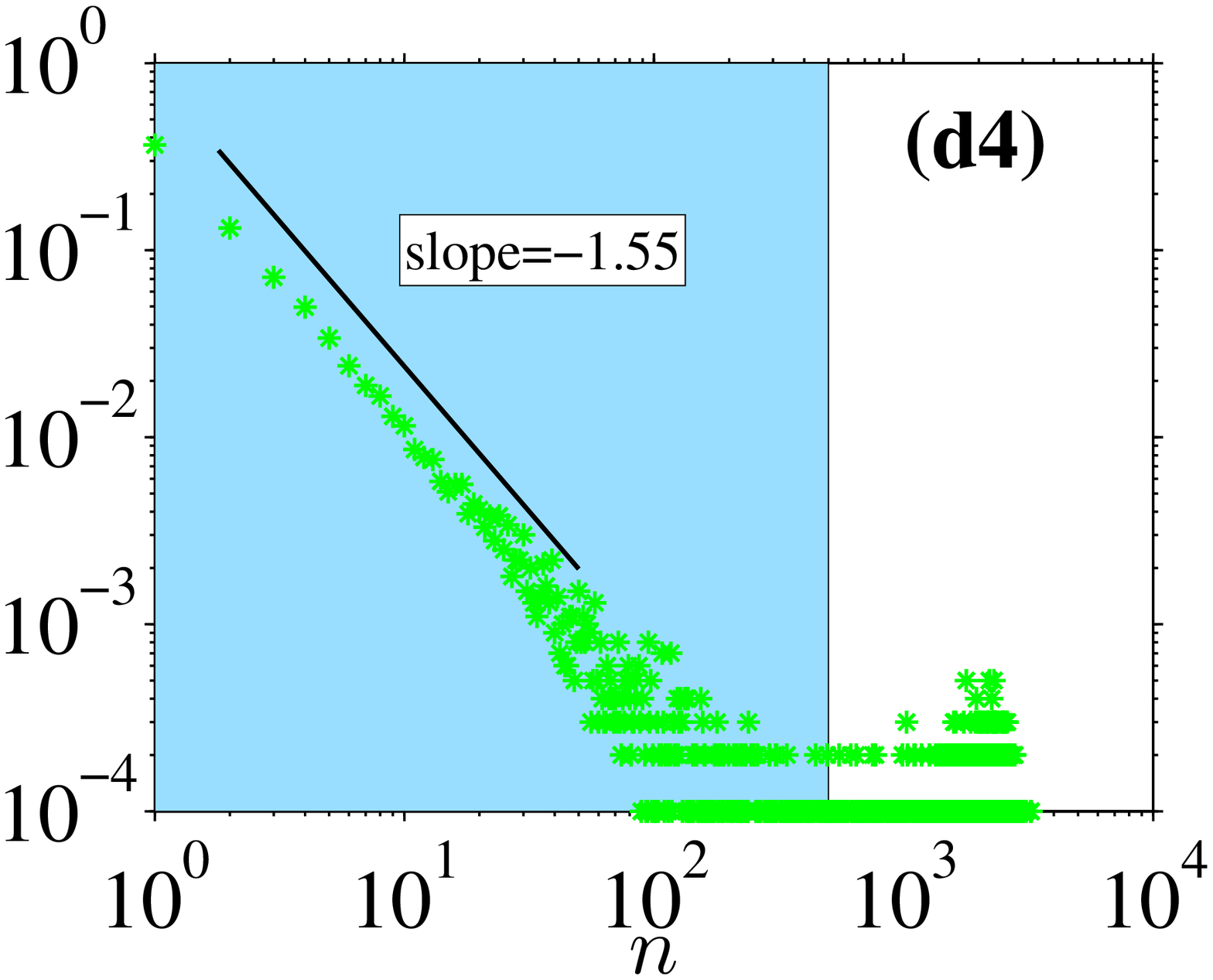}
  \caption{(Color online)\label{Fig:distribution} The cascade size
  distribution $p_c$ with randomly selecting the spreading \textit{seed} for
  different networks. The figures from left to right are simulated
  with (1) static network, (2) $\beta=0$, (3) $\beta=2$ and (4) $\beta=-2$
  respectively; from top to bottom are (a) Regular network, (b) Random
  network, (c) Small-world network and (d) Scale-free network
  respectively. The spreading rate is set to $\lambda=0.2$. The results are obtained by
  averaging over 10000 independent realizations.}
\end{figure*}

In this paper, we use the SIR model with fixed recovery time
$T=1$ to illustrate the proposed information spreading process. All
individuals in the system must be one of the three discrete states:
the uninformed individuals (defined as $S$-state), the active
informed individuals (defined as $I$-state) which would transmit information to their $S$
neighbors and the inactive informed
individuals (defined as $R$-state) which know the information but wouldn't transmit it
any more. Initially, a node is randomly selected
as the $I$-state, which is considered as the \emph{seed} for the
information spreading, and all other individuals are set as the $S$-state. At each time step,
the $I$ individuals transmit the information to the $S$ nodes through $S-I$
links with the spreading rate $\lambda$. In social network, the
individuals usually won't transfer an information item more than
once to the same neighbor, namely the non-redundancy property
 \cite{Lu-Chen-Zhou-2011-NJP}. Therefore, each $S$-$I$ link can just be
used once in our model, no matter the transmission through this link
is successful or not. That is to say, when one $I$ individual transmits
the information to all its $S$ neighbors at one time step, the $I$
individual will change to $R$-state and wouldn't be able to transmit information
any more. The information spreading process stops until
there is no $I$-state individual in the system.

Generally, information decays very fast
\cite{Wu-Huberman-2007-PNAS}, that is to say, some information would
lose attraction within a very short period. Hence, how to spread the
information quickly is a
critically important problem in the social system
\cite{Doer-Fouz-Friedrich-2012-CACM}. The link rewiring strategy is
one of the possible methods to enhance the information spreading efficiency through
changing the network structure. Consequently, we consider the link
rewiring strategy as the following way (see
Fig.~\ref{Fig:Model}). For each $S$ node (such as $j$ in Fig.~\ref{Fig:Model}) that
connects to the $I$-state nodes ($i$), $I$ node will rewire the link to a
randomly chosen $S$ node ($j'$) among the $i$'s second-order neighbors with probability $p_f$, which is
determined according to the Fermi function from statistical physics
\cite{Szabo-Toke-1998-PRE,Traulsen-Pacheco-Nowak-2007-JTB,Fu-Rosenbloom-Wang-Nowak-2011-PRSB,Traulsen-Semmann-Sommerfeld-Krambeck-Milinski-2010-PNAS}
 \begin{equation}
 \label{EQ:FermiFunction}
 p_f=\frac{1}{1+e^{-\beta(\pi_{j'}-\pi_{j})}},
 \end{equation}
 where $\pi_j,\pi_{j'}$ are respectively the number of the $S$ neighbors of two target nodes $S_j,S_{j'}$. And $\pi_{j}$
can be considered as the
payoff \cite{Traulsen-Pacheco-Nowak-2007-JTB} of strategy that
connecting with the individual $j$. The individual with
more connections (corresponding to the large payoff) is regarded
 as \emph{information hungry}, for they would
 be more likely to communicate with others, hence the information is more likely to
 spread out through such individuals. In general, individuals may know only the
 payoff with a small range instead of all social
 systems, so that the rewiring individual (node $j'$ in Fig.~\ref{Fig:Model}) is randomly chosen in the second-order
 neighbors of the corresponding $I$ individual in this model.

We assume that individuals are of finite rationality following the
common practice. That is to say, individuals prefer to choose
strategies with higher payoff, while they are also possible to select those
with lower payoff. Fermi function is a widely used way to achieve
this purpose with incorporating stochastic element into the
model. For small $\beta$,
individuals are less responsive to the payoff differences, while large
$\beta$ would weaken this stochastic effect, and individuals
may switch to the nodes with higher payoff,
even if the payoff difference is
small \cite{Traulsen-Pacheco-Nowak-2007-JTB,Fu-Rosenbloom-Wang-Nowak-2011-PRSB}.
In our model, $\beta$ is generalized to $[-\infty, +\infty]$. The
individual with larger payoff will be chosen with larger probability
when $\beta>0$, and vice verse. The rewiring probability becomes
neutral when $\beta=0$, which corresponds to the case of random
rewiring.

\section{Results \& Analysis}

The proposed model is performed on four representative networks with
the same total population $N$ and average degree $\langle k \rangle$. 1)
\textit{Regular network}: a ring lattice with $N$ nodes and $k$
edges per node \cite{Watts-Strogatz-1998-Nature}; 2)
\textit{Small-world network}: rewiring each edge at random with
probability $p_s$ based on the regular
network \cite{Watts-Strogatz-1998-Nature}; 3) \textit{Random network}:
randomly rewiring probability $p_s=1$
\cite{Watts-Strogatz-1998-Nature}; and 4) \textit{Scale-free
network}: $m=k/2$ in the BA model where $m$ is the number of edges
for the new node \cite{Barabasi-Albert-1999-Science}, and the
network exhibits a power-law degree distribution $p(k)\sim
k^{-\gamma}$ with $\gamma=3$. To alleviate the effect of randomly selecting the spreading $seed$, all the simulation results are obtained
by averaging over 10000 independent realizations. In all simulations, we set
$N=10000$ and $\langle k \rangle=6$.

\begin{figure}[htb]
  \centering
  \includegraphics[width=7.3cm]{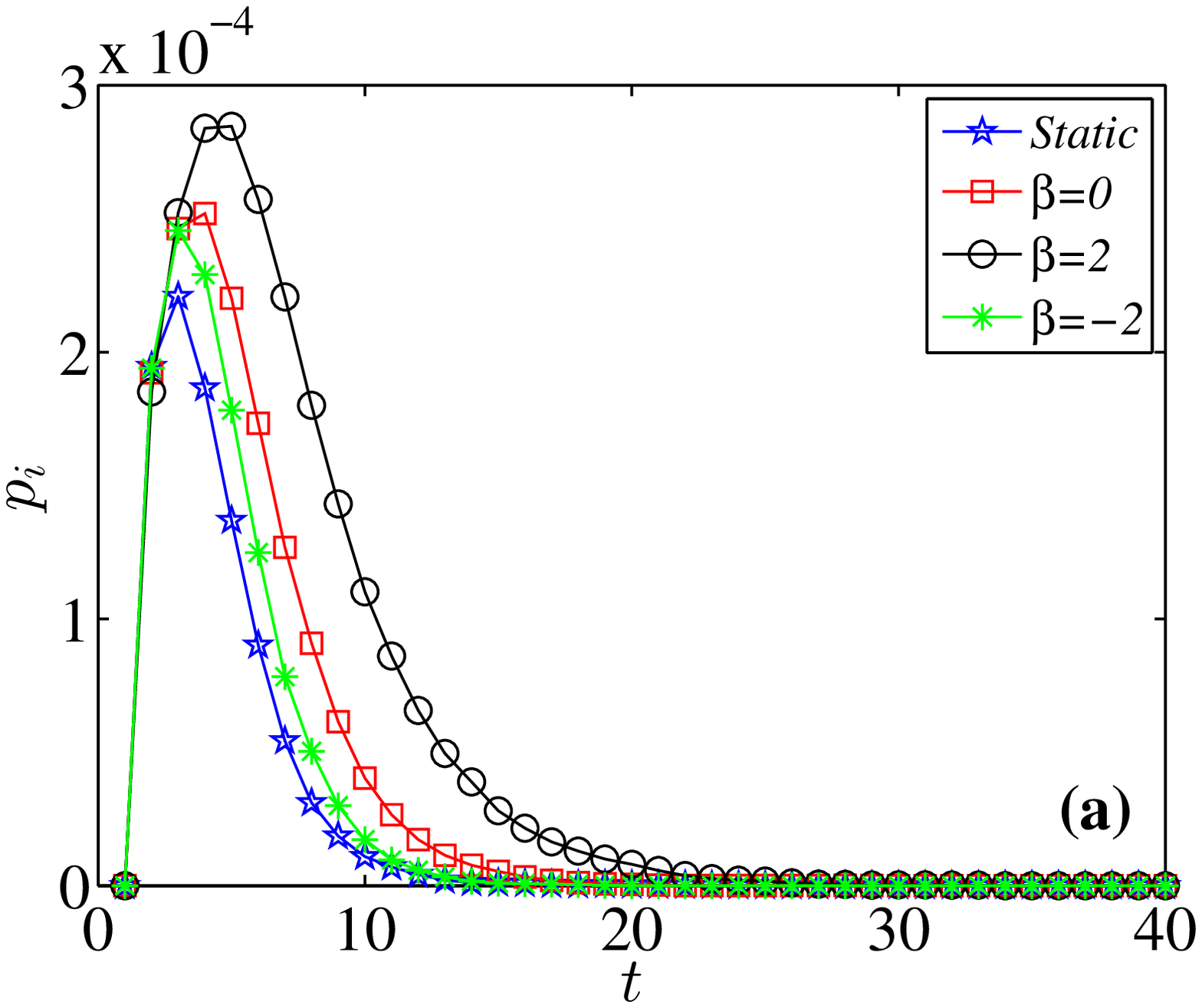}
  \includegraphics[width=7.5cm]{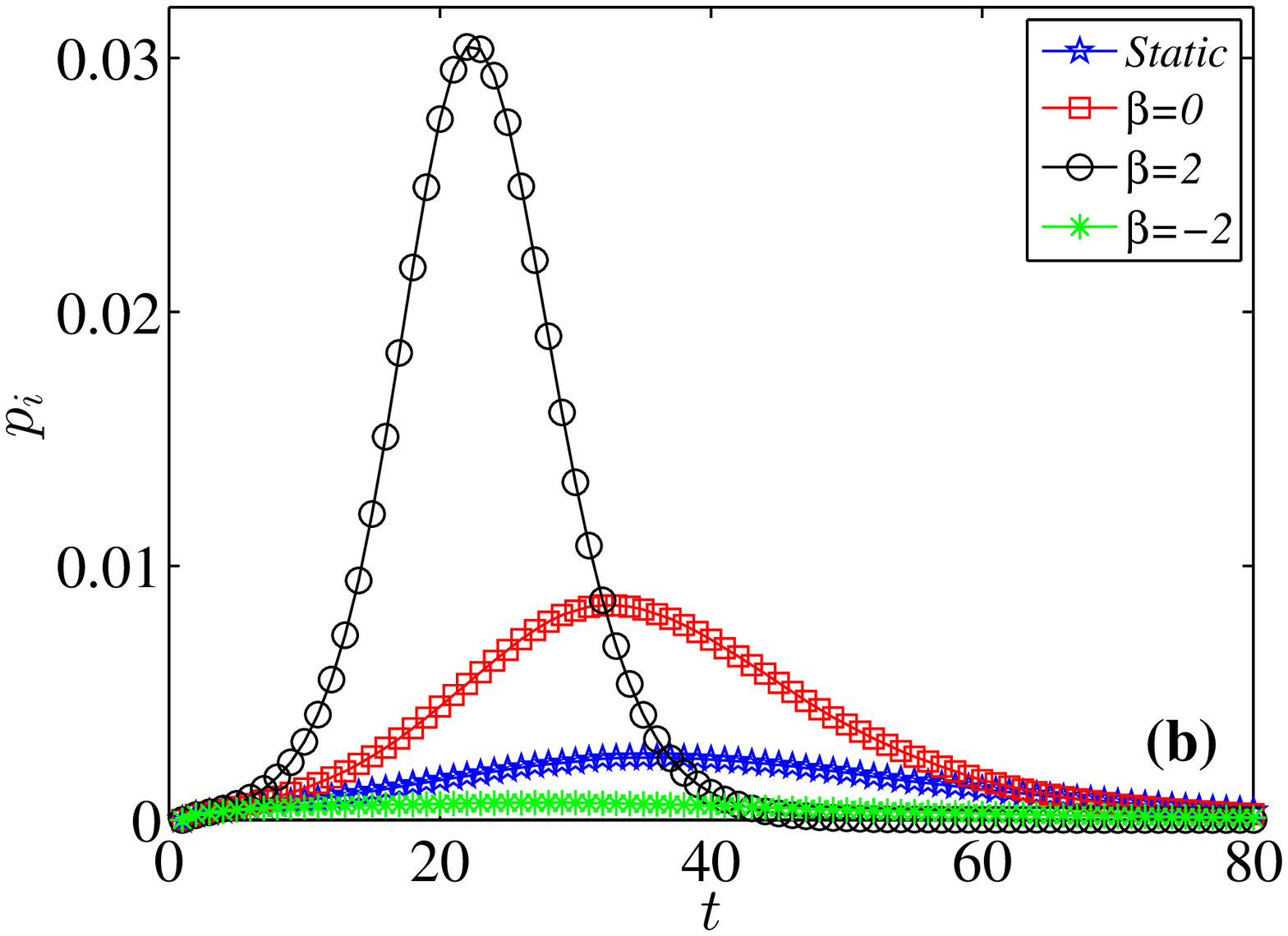}
  \includegraphics[width=7.5cm]{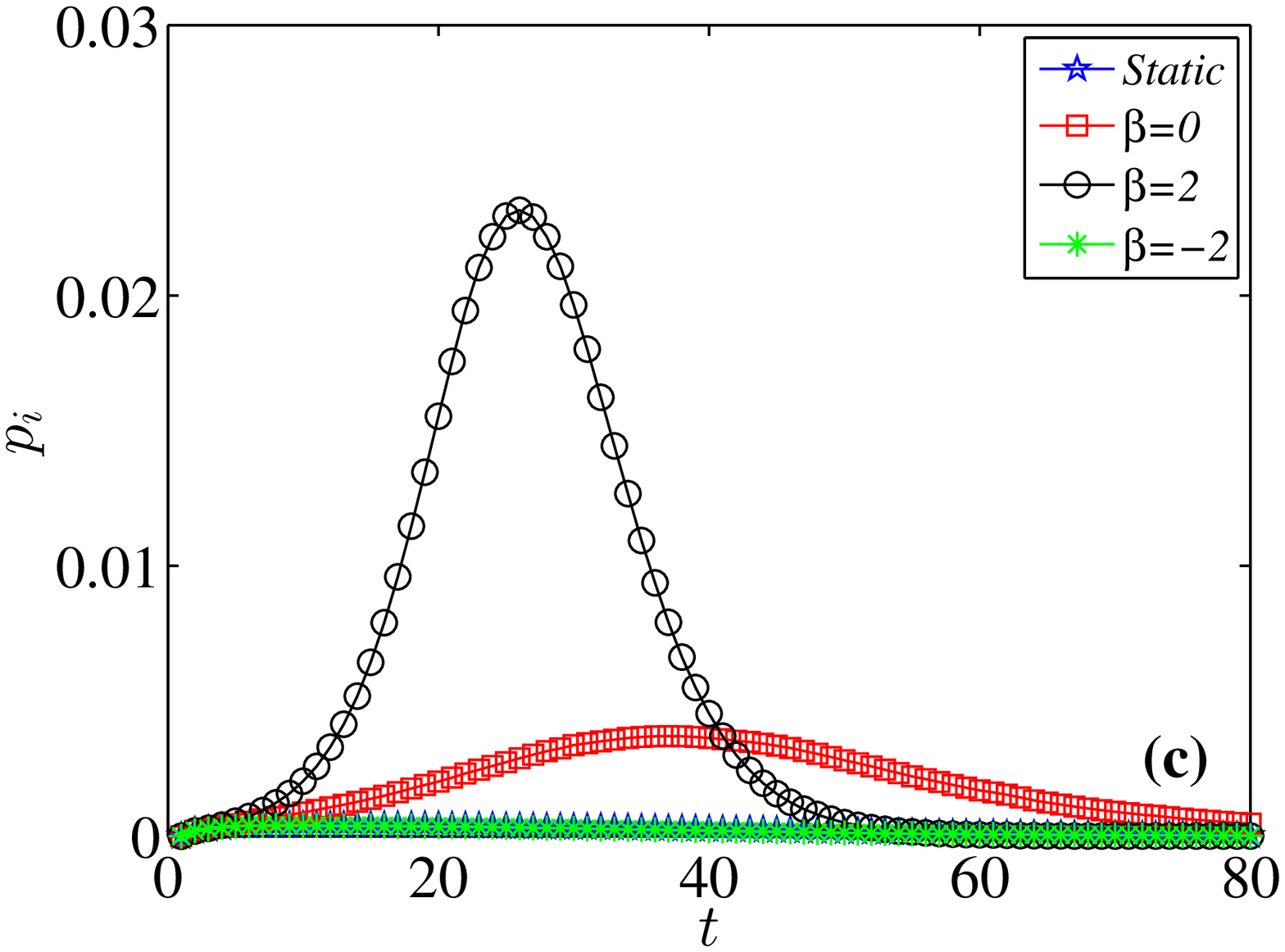}
  \includegraphics[width=7.5cm]{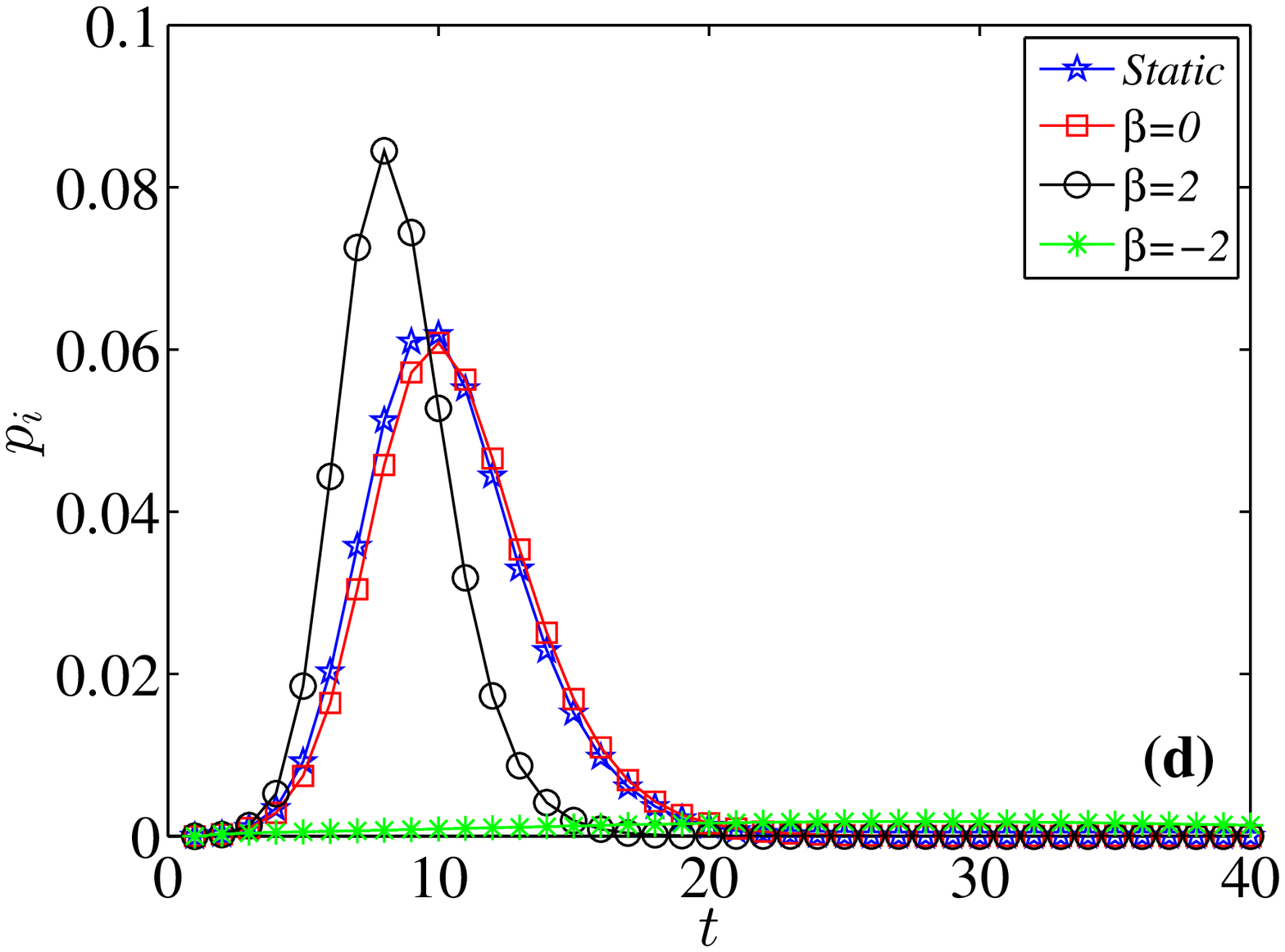}
  \caption{(Color online)\label{Fig:cascade_evolution} Dynamics of $p_i$ with different methods:
  static network (blue pentagram), $\beta=0$ (red square), $\beta=2$
   (black circle) and $\beta=-2$ (green star). All simulations are run on four representative networks: (a) Regular network; (b) Random network; (c)
  Small-world network; (d) Scale-free network. The spreading rate is set to $\lambda=0.2$. The results are obtained by averaging over $10^4$
  independent realizations.
  }
\end{figure}

To illustrate the spreading process, we firstly focus on the time evolution
of the proportion of $R$-state individuals ($p_r$)
in the systems. $p_r$ indicates the information diffusion range, and larger
$p_r$ value in the stationary state shows the broader spreading.
Fig.~\ref{Fig:P_time_lambda} illustrates the dynamics of $p_r$ with
the spreading rate $\lambda=0.2$. It can be seen that the information spreading
on scale-free network is both broader and faster than that on the
other networks. This is caused by the
heterogeneous degree distribution of the scale-free network where
the hub nodes hold the connective in the
spreading process (see Fig. \ref{Fig:Degree_time}). The number of $R$-state nodes follows the relation ``random $>$ small-world $>$
regular'' which is consistent with the traditional understanding of
the epidemic spreading \cite{Zhou-Fu-Wang-2006-PNS}. This is
different from the information spreading model considering the
social reinforcement \cite{Lu-Chen-Zhou-2011-NJP}, in which the spreading on regular network is faster and broader than
that on random network. For each kind of
network, we show the results of $\beta=2, 0, -2$ and the
static network as the baseline to illustrate the influence of the
link rewiring strategy on the information spreading process. Fig. \ref{Fig:P_time_lambda}
shows that in all observed networks, the spreading results obtained
for $\beta=2$ is broader than
others, and the enhancement is significant compared with the static
network. $\beta>0$ indicates the case that the $I$-state nodes are more
likely to rewire the links to nodes with more $S$ neighbors, and the
informed chance of the corresponding $S$ nodes will increase through the
rewired links. Once the information is transformed to the large payoff $S$ individuals,
it will spread out quickly for more $S$ neighbors will be informed through the
corresponding individuals at the next step. In contrast, the spreading process will \textit{die}
quickly for $\beta<0$. Furthermore, it is interesting to find out that the influence
of the link rewiring strategy on the scale-free network is also
quite different from others. Neutrally rewiring strategy ($\beta=0$)
enhances the spreading on regular, small-world and random networks,
while slows the diffusion on scale-free networks, since it breaks
the already present preferential attachment. According to Fig. \ref{Fig:P_time_lambda}(a), it is obtained
that the enhancement of the rewiring strategy on the regular network is very weak.
With considering the limited attention \cite{Weng-Flammini-Vespignani-Menczer-2012-SR}
of the individuals in social systems, the present rewiring range is confined in
the second-order neighbors of the corresponding $I$-state individual. But the
second-order neighbors are also just the neighbors with large probability in the regular network for the large
clustering coefficient. As a result, the information spreading enhancement is very limited with the
proposed rewiring strategy for the regular network.

\begin{figure}[htb]
  \centering
  \includegraphics[width=7.5cm]{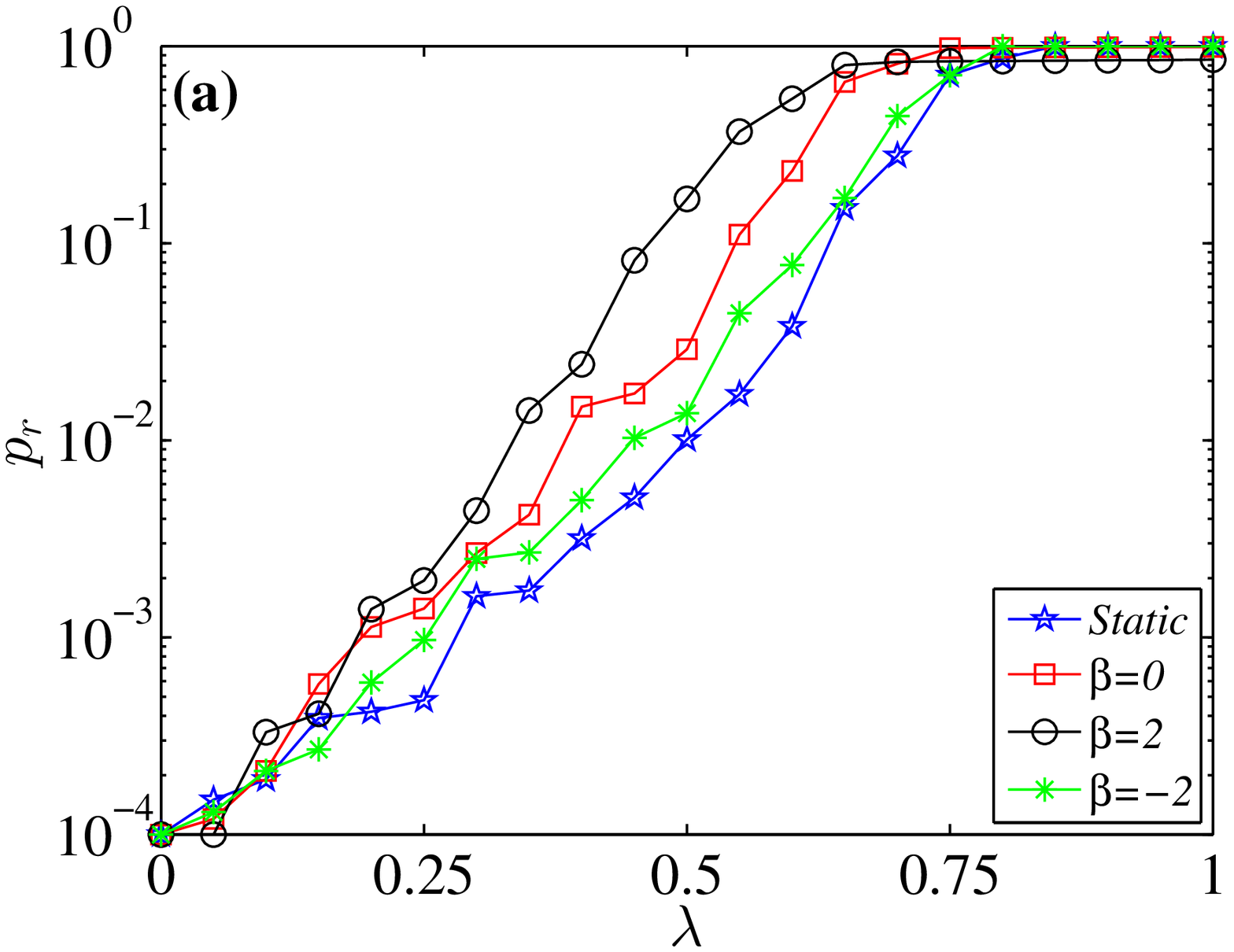}
  \includegraphics[width=7.5cm]{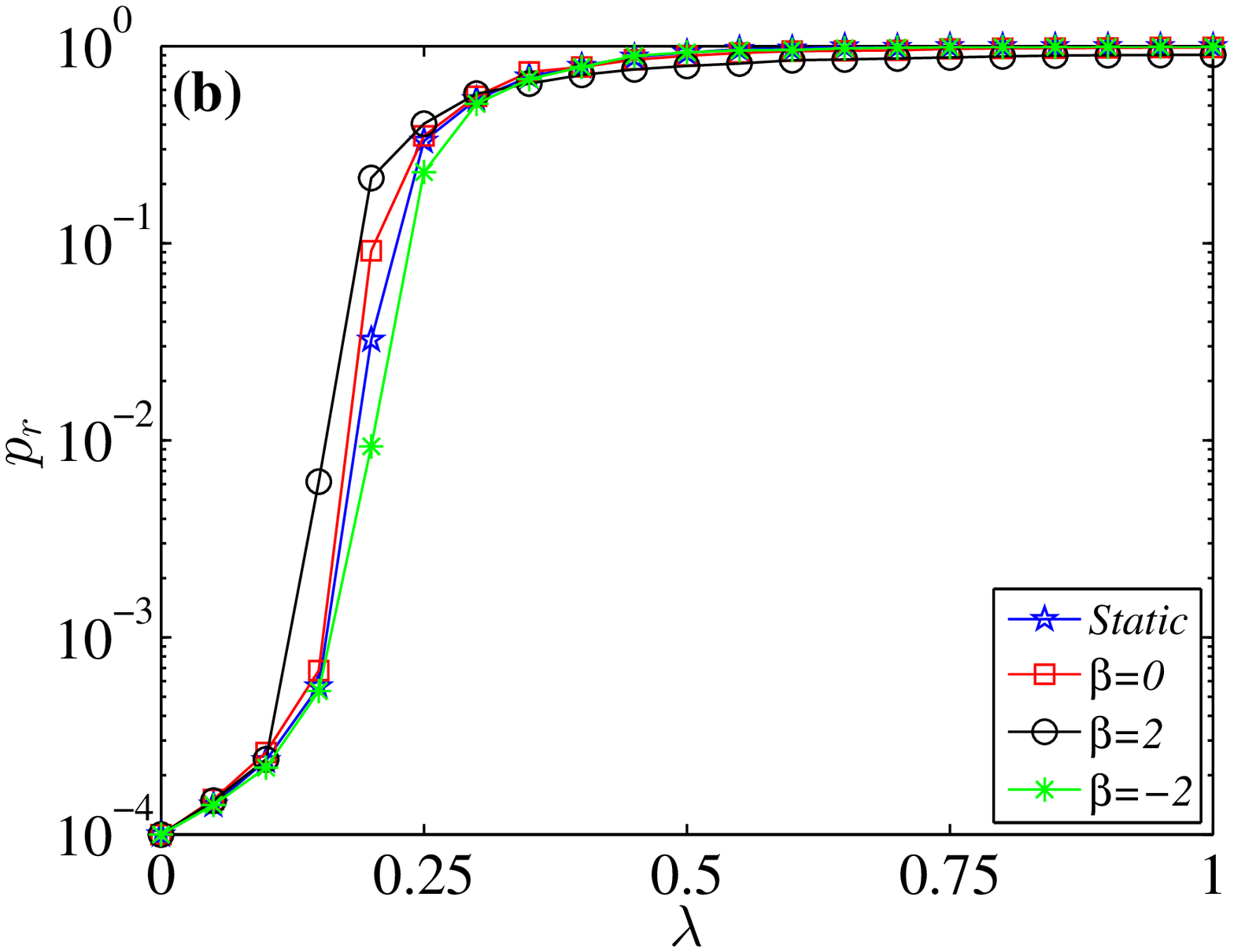}
  \includegraphics[width=7.5cm]{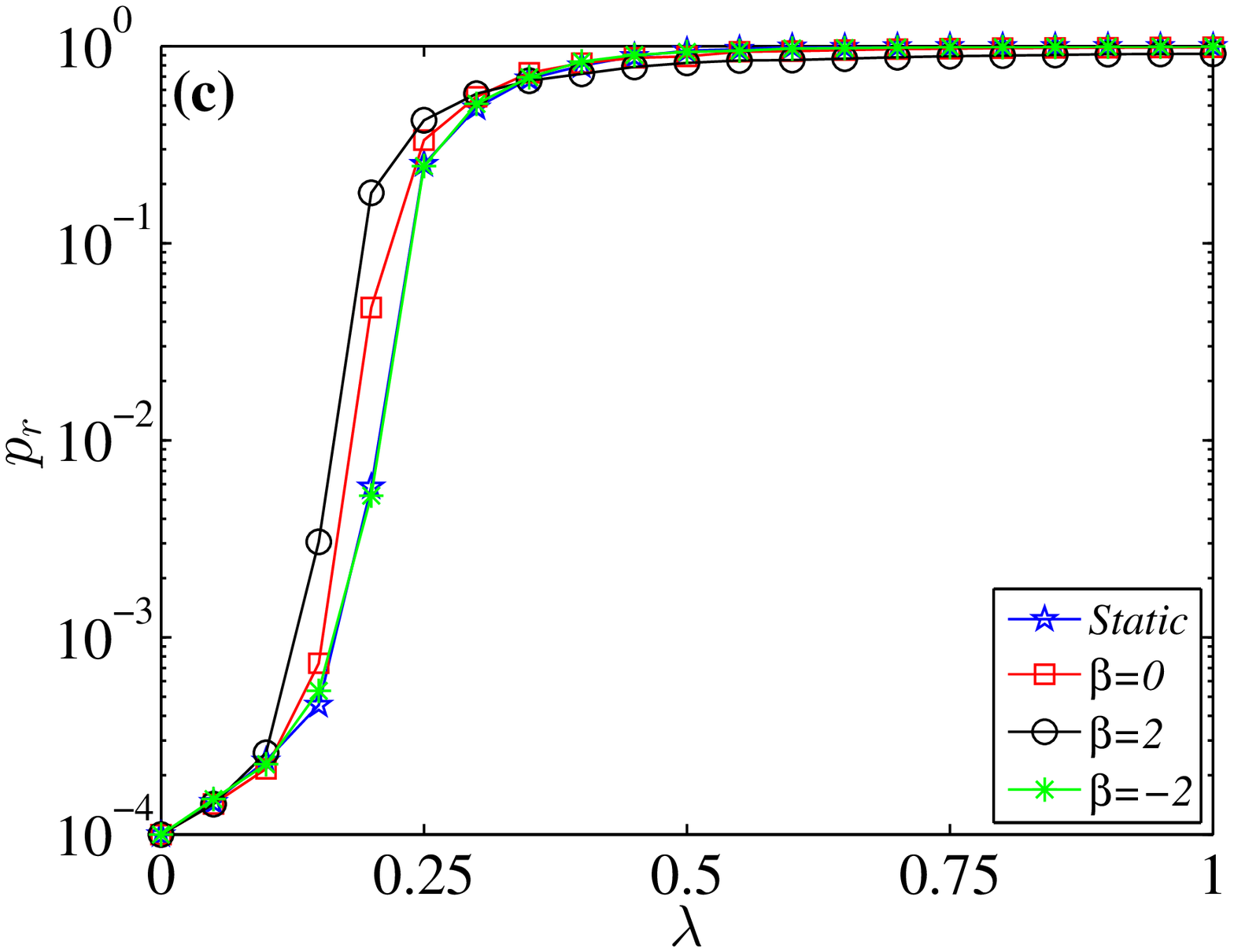}
  \includegraphics[width=7.5cm]{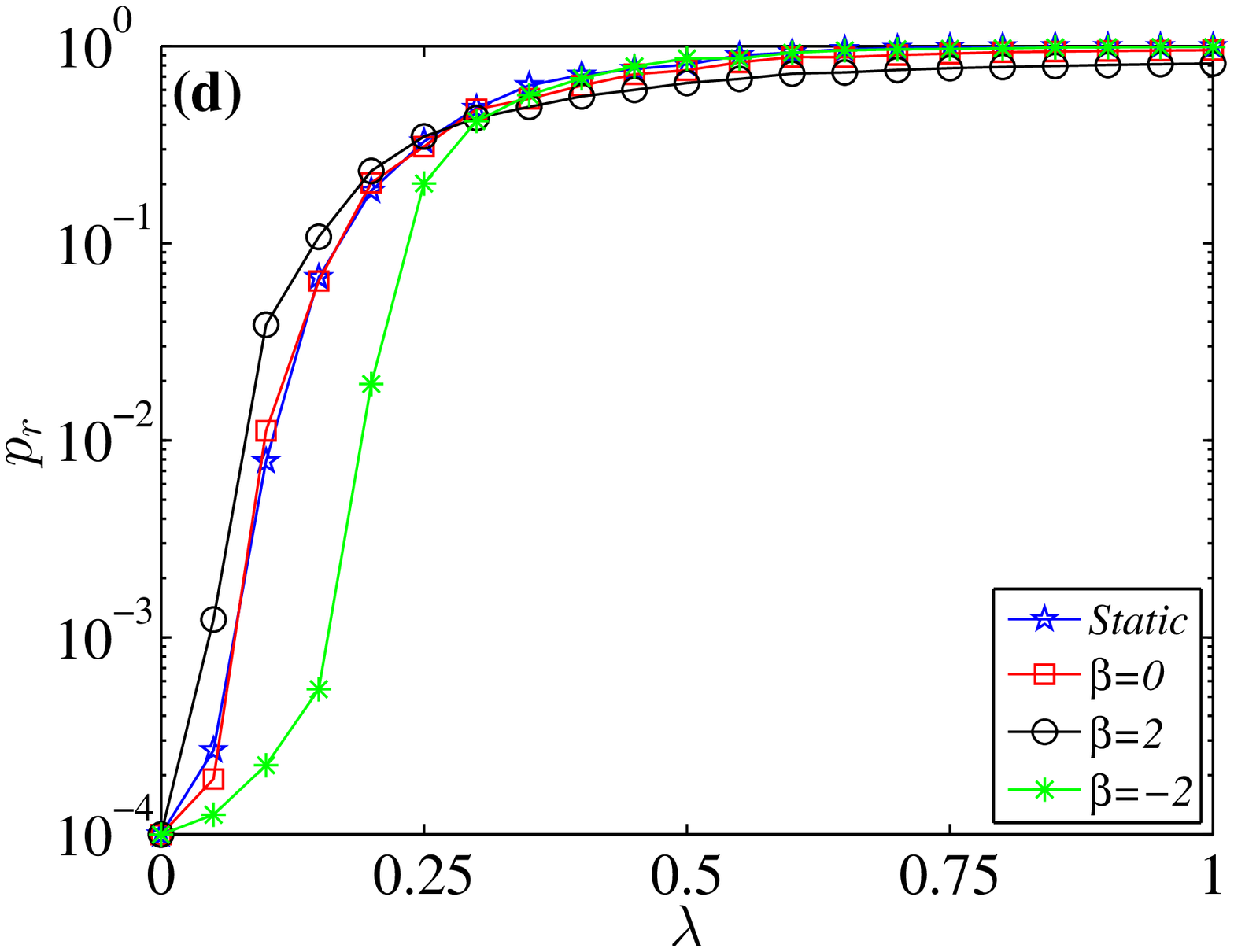}
  \caption{(Color online)\label{Fig:SizeVSLambda} The dependence of $p_r$ at the final state on the spreading rate $\lambda$
  with different methods: static network (blue pentagram), $\beta=0$ (red square), $\beta=2$
   (black circle) and $\beta=-2$ (green star). All simulations are run on four representative networks: (a) Regular network,
  (b) Random network, (c) Small-world network and (d) Scale-free network. }
\end{figure}

The informed individuals in the network can be considered as the
\emph{cascade} \cite{Fowler-Christakis-2010-PNAS}, which is a set of
the informed individuals generated by the information spreading
process. Fig.~\ref{Fig:distribution} shows the cascade
size distribution (CSD) over 10000 independent realizations. The
CSD can be considered as two regimes: a power-law for
small sizes (the blue area in Fig. \ref{Fig:distribution}), and the rest part for large sizes which is very different
over the network structure and parameter $\beta$. According to
Fig.~\ref{Fig:distribution}, it can be found that the stronger
heterogeneity of the network is, the more
sharply the power-law range decays and the more higher peaks the
large size part exhibits. The CSD for large size ($n >$10)
on the scale-free network (the bottom four subgraphs which are marked with $d$)
can be described as a log-normal function, which is very similar
to the empirical results on \emph{Digg}\footnote{http://digg.com/}
where the information spreading on the fans' network with the power-law degree distribution structure
\cite{VerSteeg-Ghosh-Lerman-2011-ICWSM}. For the same network
structure, there are more cascades with large size for $\beta>0$, resulting in more informed individuals in
the stationary state which coincides with
Fig.~\ref{Fig:P_time_lambda}. It is interesting to find out that the
two regimes of the CSD are separated absolutely
in the scale-free network for $\beta=2$
(Fig.~\ref{Fig:distribution}(d3)) and there is nearly no cascade
with the size ranging from 10 to 4000, which indicates that the
information will either spread to a high level or
die quickly. Therefore, the spreading at the initial steps are very
important. If the information is still survival after the
beginning several steps, it will spread into a fraction of the
total population. In order to illustrate the evolution of the
information cascade, we investigate the dynamics of $I$-state individuals proportion
$p_i$ in Fig.~\ref{Fig:cascade_evolution}. In this model, the $I$-state individuals are
just the newly informed individuals in the last step. For all the networks, the $p_i$ increases sharply at
several initial steps, then decrease until the spread process stops.
The $p_i$ with positive $\beta$ is much larger than the
corresponding static networks, which means that the information
spreads much faster with the link rewiring strategy. In the
scale-free networks, the spreading is quite quick that more than 8\% of the total population are informed in a single step.

Fig.~\ref{Fig:SizeVSLambda} displays the final spreading level for each network structure
 as a function of the spreading rate $\lambda$. It is easy to realize that
$p_r$ increases with $\lambda$ for all the rewiring strategies and network structures.
When $\lambda$ is small, the information spreading
enhancement with the positive $\beta$ could be very significant. In addition, the curve with positive
$\beta$ shows that the $p_r$ peaks suddenly with small $\lambda$, which
means that positive $\beta$ makes the critical
$\lambda$ value diminish. For large $\lambda$, nearly all the individuals will
be changed to the $R$-state. Furthermore, it is very obvious that there is the inhibitory
effect on information spreading using the rewiring strategy with negative $\beta$ on the scale-free
network according to Fig.~\ref{Fig:SizeVSLambda}(d).


\begin{figure}[htb]
  \centering
  \includegraphics[width=8.5cm]{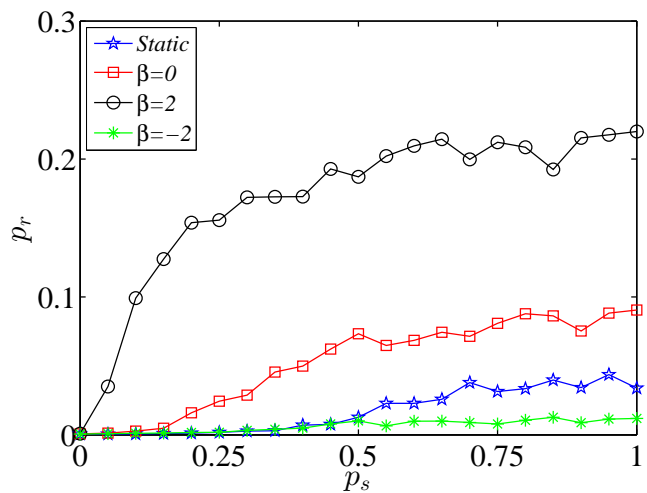}
  \caption{(Color online)\label{Fig:P_Ps} $p_r$ at the final state as a function of parameter $p_s$ with different
  methods: static network (blue pentagram), $\beta=0$ (red square), $\beta=2$
  (black circle) and $\beta=-2$ (green star). The spreading rate is set to $\lambda=0.2$.}
\end{figure}

 The aforementioned results show that spreading process on regular network and random network
are quite different. The network structure could be parameterized by the
randomly rewiring probability $p_s$ (namely 'small-world' parameter), where $p_s=0$ and $p_s=1$
correspond to regular and random network, respectively. We plot $p_r$ at the stationary state as a
function of $p_s$ in Fig.~\ref{Fig:P_Ps}. Similar to the previous
studies \cite{Zanette-2001-PRE,Kuperman-Abramson-2001-PRL}, we also
find out that $p_r$
increases with the small-world parameter $p_s$. In the small-world
network, as the existence of the long range connection, the average
distance of the network will
decrease \cite{Watts-Strogatz-1998-Nature}. The information spreading
should be faster and broader, as the distance between $I$
and $S$ nodes becomes shorter. The influence of the link rewiring with Fermi
function is robust with the change of $p_s$, and the information
spreading is much broader for $\beta>0$. Experiments
show that, with positive $\beta$, even very small $p_s$ can bring
a remarkable improvement in promoting $p_r$.

\begin{figure}[htb]
 \centering
 \includegraphics[width=8.5cm]{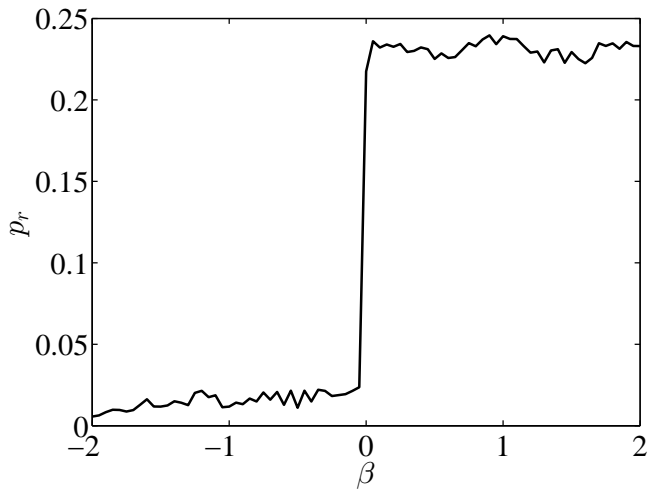}
 \caption{(Color online)\label{Fig:P_Beta} $p_r$ at the final state as a
 function of the parameter $\beta$ on scale-free network. The spreading rate is set to $\lambda=0.2$.}
 \end{figure}

In order to illustrate how the parameter $\beta$ influences the
spreading process, we plot $p_r$ as a function of parameter $\beta$ on the scale-free network in
Fig.~\ref{Fig:P_Beta}. The simulation result shows that a transition
occurs between a regime where the spreading process dies out within a
small neighborhood of the \emph{seed}, and a regime where it spreads
over a finite fraction of the whole population. The spreading is very
limited ($p_r$ is round 1\%) for $\beta<0$, where the number of $R$-state
individuals raises very slightly with increasement of $\beta$. It means
that rewiring active links to the nodes with less $S$
neighbors will prevent information spreading. And this could
be used as a strategy to control the spreading for virus and rumors.
The spreading broadens very sharply when $\beta$ increases around 0,
which indicates that $\beta=0$ should be the saltation point. For
$\beta>0$, the active link will be more likely to rewire to the
nodes with more $S$ neighbors. In addition, as long as $\beta$ is positive, the
information will spread into a very broad range ($p_r$ around 23\%). The behavior of the informed numbers around
$\beta=0$ indicates that, the symbol of $\beta$ rather than the accurate
numerical value is the most significant factor to affect the
information spreading with the proposed rewiring strategy.

\section{Discussion}

\begin{figure}[htb]
  \centering
  \includegraphics[width=7.5cm]{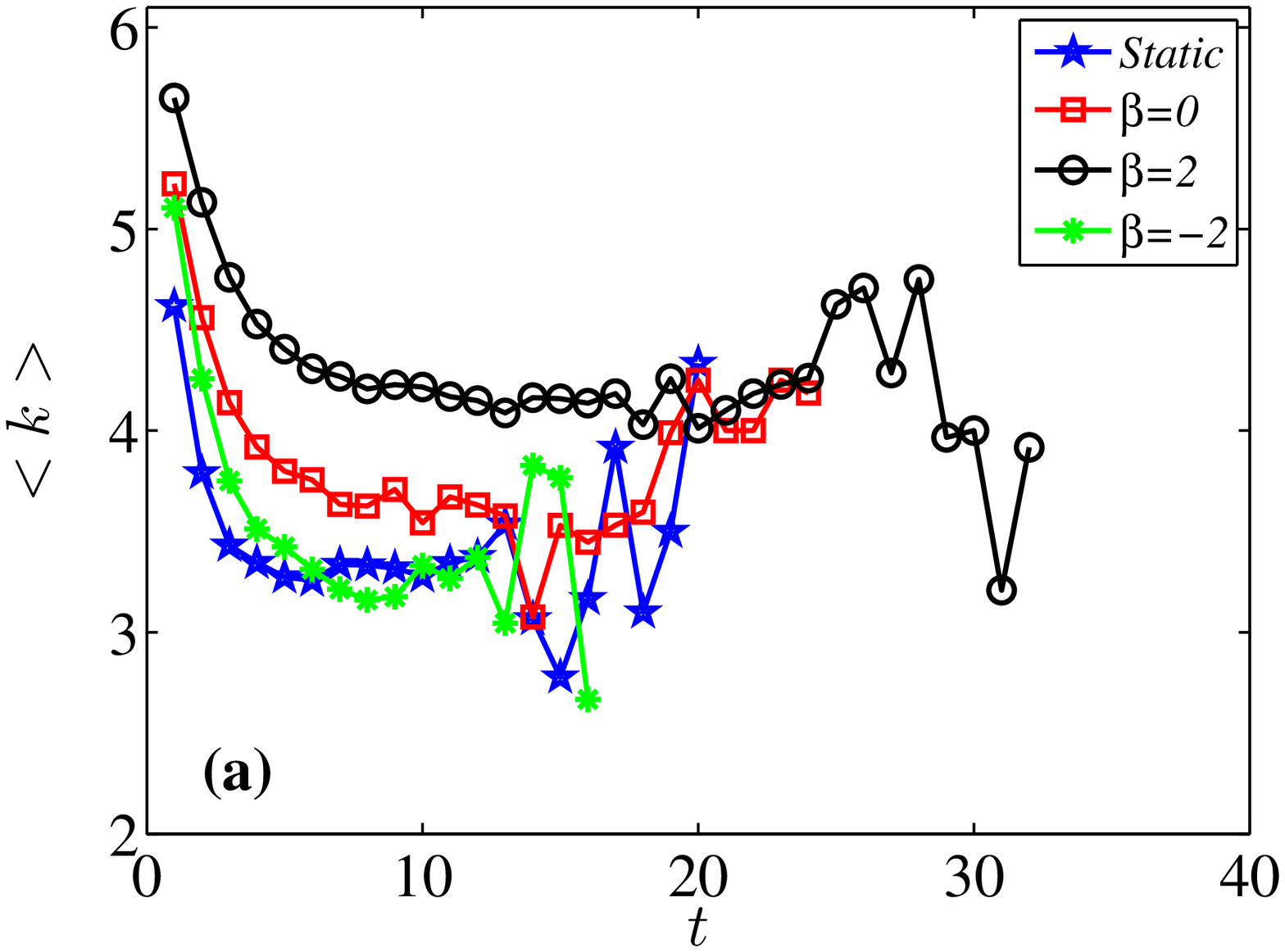}
  \includegraphics[width=7.5cm]{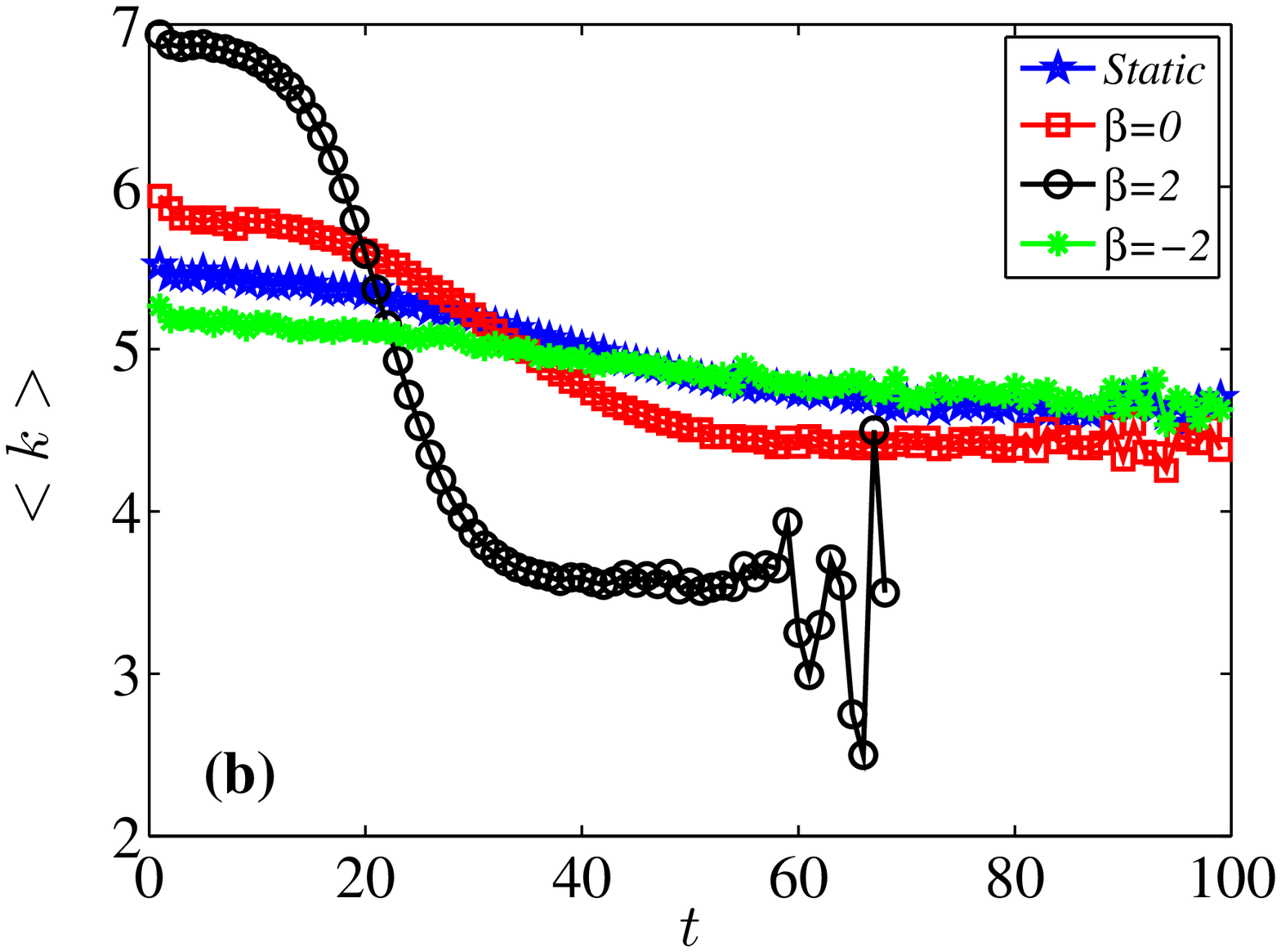}
  \includegraphics[width=7.5cm]{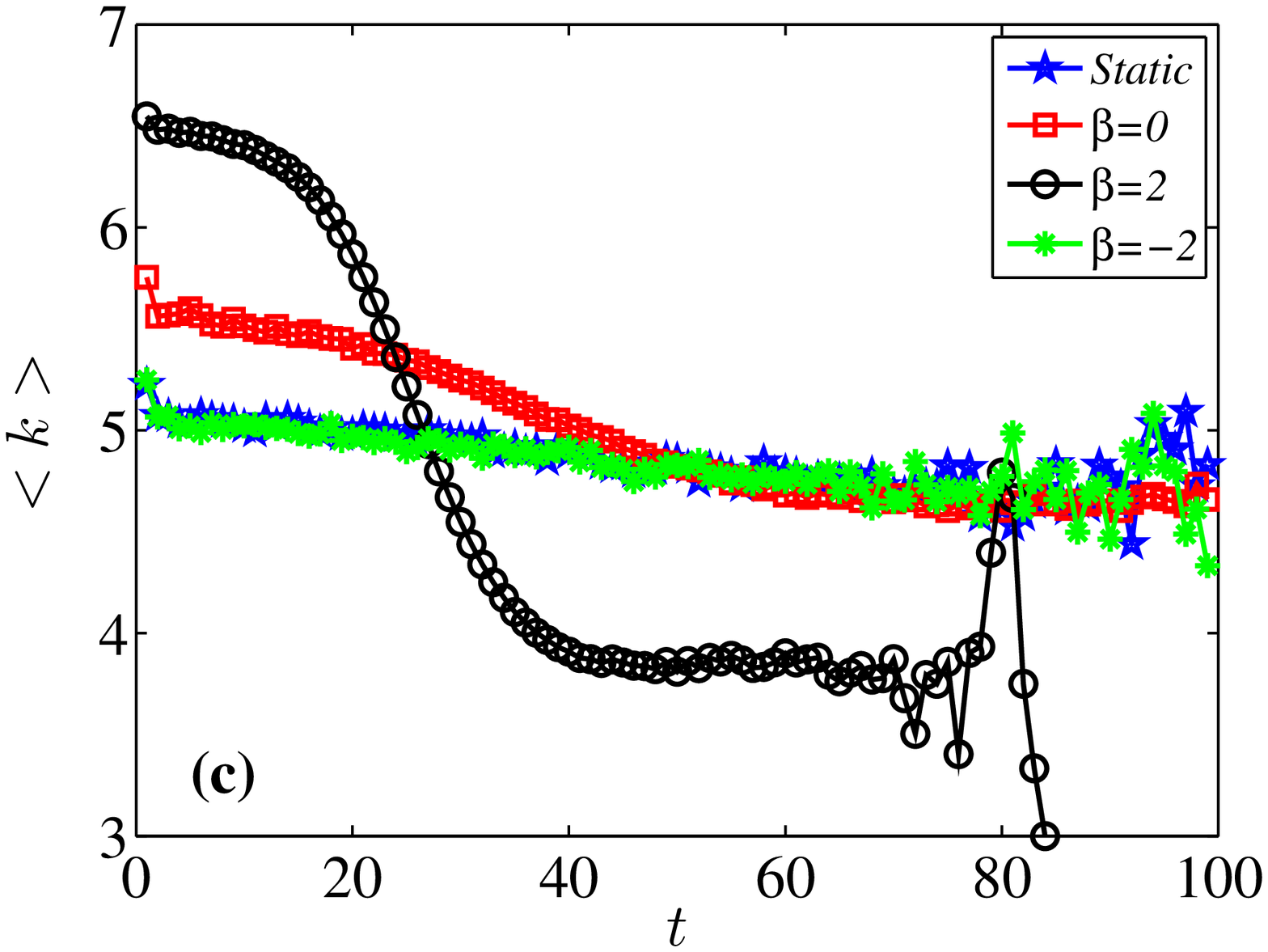}
  \includegraphics[width=7.5cm]{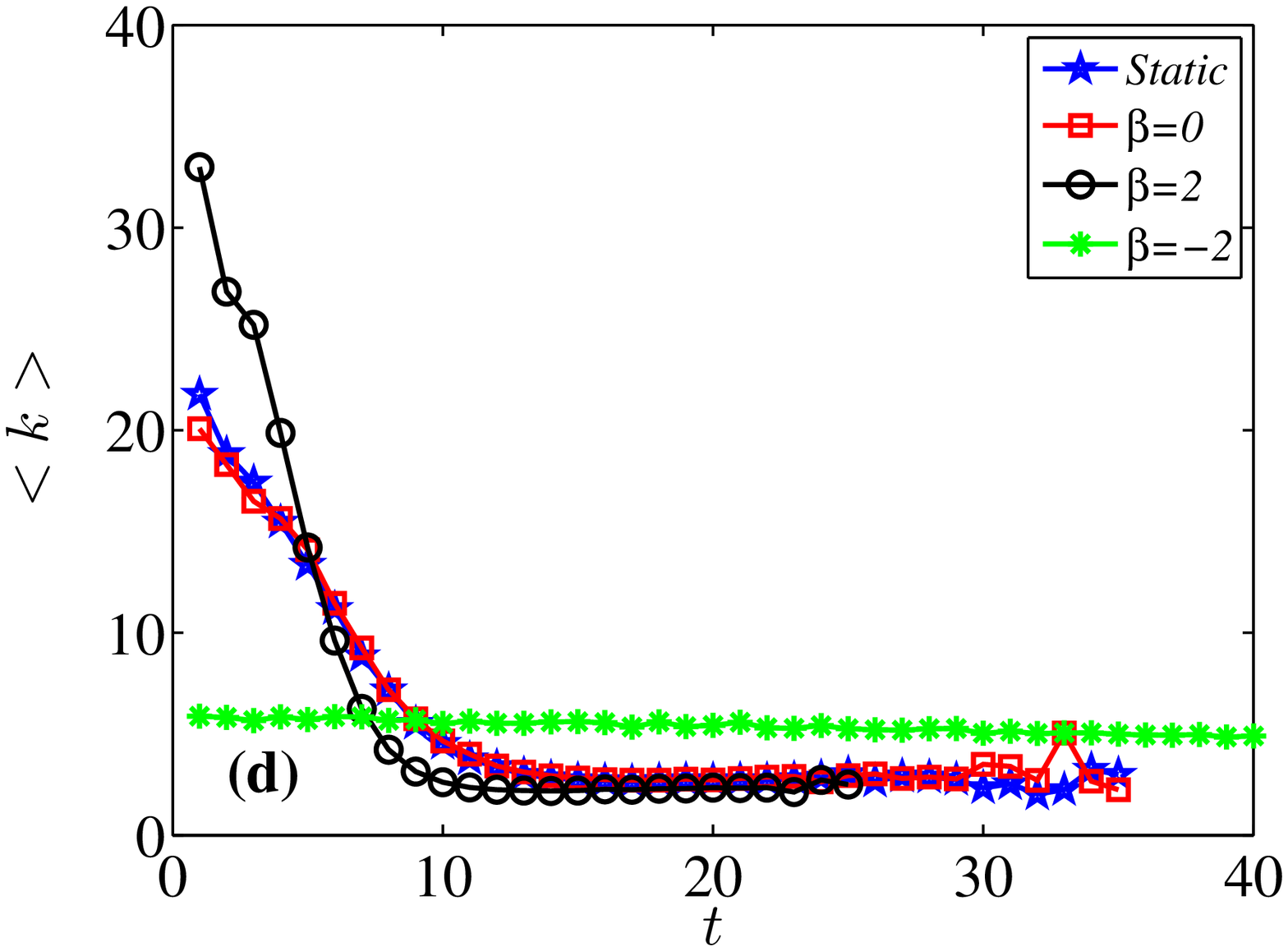}
  \caption{(Color online)\label{Fig:Degree_time} The average number ($\langle k \rangle$) of the $S$ neighbors
  of the $I$-state individuals at each time step with different methods: static network (blue pentagram), $\beta=0$ (red square), $\beta=2$
   (black circle) and $\beta=-2$ (green star). All simulations are run on four representative networks: (a) Regular network; (b) Random
  network; (c) Small-world network; (d) Scale-free network. The spreading rate is set to $\lambda=0.2$. The results are obtained by averaging over $10^4$
  independent realizations.}
\end{figure}

The simulation results show that the positive $\beta$ induces a
broader and faster information spreading process. To interpret the reason why
the positive $\beta$ enhances the spreading, we investigate the dynamics
of the average number of $S$ nodes connecting with the newly
informed nodes in Fig.~\ref{Fig:Degree_time}. The results
of the four curves at the beginning
of the spreading process coincide with Fig.
\ref{Fig:P_time_lambda}, which suggests that the information item
will spread broader and faster if the nodes with more $S$ neighbors
are informed as the spreading initializes. In addition, the result on the
scale-free network is consistent with Barth\'elemy's work
\cite{Barth-Barrat-Pastor-Satorras-Vespignani-2004-PRL}, where the
dynamics of the spreading is characterized by a hierarchical
structure, that the information is transformed to large degree nodes
firstly, then pervades the network via smaller degree nodes
rapidly. And the hierarchical spreading patten would be more obvious
when $\beta>0$ (see Fig.~\ref{Fig:Degree_time}(d)).
However, the large-degree nodes don't always speed up the
spreading process, such as the game-theoretic models of the
innovation spread \cite{Montanari-Saberi-2010-PNAS}.

For the positive $\beta$ on the generalized Fermi function, we will
obtain large rewiring probability by $\frac{\displaystyle
1}{\displaystyle 1+e^{-\beta(\pi_{j'}-\pi_{j})}}$ if the payoff of
$S$-state node $j'$ is larger than $j$, and vice verse. That is to
say, the $I$ nodes will be more likely to rewire the link to the $S$
nodes with more $S$ neighbors. Following this strategy, the nodes
with more $S$ neighbors have more chance to be informed, and
the information could be more likely to spread out through such
nodes. Therefore, we can observe more $S$ neighbors of the
newly informed individuals with positive $\beta$ (Fig.
\ref{Fig:Degree_time}). For $\beta<0$, the $I$ individuals will
more likely reconnect to the $S$ individuals with less $S$
neighbors, leading to quick annihilation of the spreading.

\section{Conclusion}

In this paper, we proposed a dynamic model for the information
spreading process that considers the link rewiring based on the SIR
model with fixed recovery time $T=1$. The rewiring probability is
chosen following the generalized Fermi function based on the payoff
between the two selected uninformed individuals. Simulation
results on four representative networks show that the information will spread broader and faster
when the parameter $\beta>0$, because that the uninformed individuals
with more uninformed neighbors are more likely to be informed at the
beginning spreading steps. Through those
uninformed hubs, the information item can spread into a finite
proportion of the population quickly. The cascade size distribution
indicates that the initial steps are very important in the information spreading process,
where the information can spread into a finite fraction if it can survive at the beginning several
steps. In addition, the negative $\beta$ can be used as a strategy
to control the spreading of virus and rumors.

Recently, the research of the information spreading based on
temporal networks has attracted more and more attention
 \cite{Karsai-Kivela-Pan-Kaski-Kertesz-Barabasi-Saramaki-2011-PRE,Iribarren-Moro-2011-PRE}.
Simulation results in this paper demonstrated that with the large
payoff trend, the rewiring strategy can significantly enhance the information
spreading process. Moreover, it is found that the human
communication pattern is of critical importance in the information
diffusion \cite{Miritello-Moro-Lara-2011-PRE}, which provides a promising way to enhance the efficiency of
\emph{Information Filtering} \cite{Lv2012, qiu2013PO, qiu2011EPL} in the era of big data. For a more detailed
evaluation, the temporal patterns of human communication such as the
burst activity should be studied on the rewiring strategy in future
work.

\section*{Acknowledgments}

This work was partially supported by Natural Science Foundation of
China (Grant Nos. 11105024, 11205040, 1147015 and 11301490), the EU FP7 Grant 611272 (project GROWTHCOM), Zhejiang Provincial Natural Science
Foundation of China (Grant No. LY12A05003 and LQ13F030015), the start-up foundation and Pandeng project of Hangzhou Normal University.

\section*{References}
\bibliographystyle{elsarticle-num}
\bibliography{Bibliography}

\end{document}